\documentclass[%
twocolumn,
superscriptaddress,
nofootinbib,
 amsmath,amssymb,
 aps, prd
]{revtex4-1}

\usepackage{graphicx}
\usepackage{tensor}
\usepackage{siunitx}
\usepackage{xcolor} 
\usepackage[normalem]{ulem} 
\usepackage{lipsum} 

\usepackage[pdfborder={0 0 0}, plainpages=false, hyperfigures=true]{hyperref}

\newcommand{\beq}{\begin{equation}}
\newcommand{\eeq}{\end{equation}}
\newcommand{\beqn}{\begin{eqnarray}}
\newcommand{\eeqn}{\end{eqnarray}}

\newcommand{\lo}{\mathrel{\raise.3ex\hbox{$<$}\mkern-14mu
    \lower0.6ex\hbox{$\sim$}}}
\newcommand{\go}{\mathrel{\raise.3ex\hbox{$>$}\mkern-14mu
    \lower0.6ex\hbox{$\sim$}}}

\newcommand{\WSU}{\affiliation{Department of Physics \& Astronomy,
	Washington State University, Pullman, Washington 99164, USA}}
\newcommand{\UNH}{\affiliation{Department of Physics, University of New Hampshire, 9 Library Way, Durham NH 03824, USA}}
\newcommand{\TAPIR}{\affiliation{TAPIR, Walter Burke Institute for Theoretical Physics, MC 350-17, California Institute of Technology, Pasadena, California 91125, USA}}
\newcommand{\Cornell}{\affiliation{Cornell Center for Astrophysics and Planetary Science, Cornell University, Ithaca, New York, 14853, USA}}
\newcommand{\MPI}{\affiliation{Max-Planck-Institut fur Gravitationsphysik, Albert-Einstein-Institut, D-14476 Potsdam, Germany}}

\begin{document}
\title{High-accuracy waveforms for black hole-neutron star systems with spinning black holes}
\author{Francois Foucart} \UNH
\author{Alexander Chernoglazov}\UNH
\author{Michael Boyle}\Cornell
\author{Tanja Hinderer}\affiliation{GRAPPA Institute of High-Energy Physics, University of Amsterdam, Science Park 904, 1098 XH Amsterdam, Netherlands}
\affiliation{Delta Institute for Theoretical Physics, Science Park 904, 1090 GL Amsterdam, The Netherlands}\affiliation{Institute for Theoretical Physics, Utrecht University, Princetonplein 5, 3584 CC Utrecht, The Netherlands}
\author{Max Miller}\UNH
\author{Jordan Moxon}\TAPIR
\author{Mark A. Scheel} \TAPIR
\author{Nils Deppe}\TAPIR
\author{Matthew D. Duez} \WSU
\author{Francois H\'{e}bert} \TAPIR
\author{Lawrence E. Kidder} \Cornell
\author{William Throwe}\Cornell
\author{Harald P. Pfeiffer} \MPI


\begin{abstract}
The availability of accurate numerical waveforms is an important requirement for the creation and calibration of reliable waveform models for gravitational wave astrophysics.
For black hole-neutron star binaries, very few accurate waveforms are however publicly available. Most recent models are calibrated to a large number of older simulations
with good parameter space coverage for low-spin non-precessing binaries but limited accuracy, and a much smaller number of longer, more recent simulations limited to non-spinning black holes.
 In this paper, we present long, accurate numerical waveforms for three new systems that include rapidly spinning black holes, and one precessing configuration. We study in detail the accuracy
 of the simulations, and in particular perform for the first time in the context of BHNS binaries a detailed comparison of waveform extrapolation methods to the results of Cauchy Characteristic Extraction.
 The new waveforms have $<0.1\,{\rm rad}$ phase errors during inspiral, rising to $\sim (0.2-0.4)\,{\rm rad}$ errors at merger, and $\lesssim 1\%$ error in their amplitude. We compute the faithfulness of 
recent analytical models to these numerical results, and find that models specifically designed for BHNS binaries perform well ($F>0.99$) for binaries seen face-on. For edge-on observations, particularly for
precessing systems, disagreements between models and simulations increase, and models that include precession and/or higher-order modes start to perform better than BHNS models that currently lack these
features.
\end{abstract}

\maketitle

\section{Introduction}

Over the last 5 years, our ability to observe gravitational waves from merging compact objects has grown at an impressive rate.
The first observation of two merging black holes (GW150914)~\cite{Abbott:2016blz} was followed by nine more black hole mergers during the
``O1'' and ``O2'' observing runs of the LIGO-Virgo Collaboration (LVC)~\cite{LIGOScientific:2018mvr}, as well as the first detection of a binary neutron
star merger (GW170817)~\cite{2017ApJ...848L..13A}. The O3 observing run is still being analyzed by the LVC, but public alerts indicate that dozens
of additional mergers have been observed.\footnote{See e.g. https://gracedb.ligo.org/superevents/public/O3/}

The third type of binary merger likely to be detected by current observatories, black hole-neutron star (BHNS) mergers, remains the most elusive.
At the time of this writing, five O3 alerts are classified as likely BHNS mergers. However, the robust classification of a specific event as a
BHNS binary remains a difficult task, due to uncertainties about the mass range of black holes and neutron stars, and the inability
of current observatories to directly demonstrate the presence of a neutron star in the absence of an electromagnetic counterpart to
the gravitational wave signal. The classification of an event as a BHNS merger in public alerts only indicates the likely presence
of an object of mass $M<3M_\odot$, that could also be a low-mass black hole. Most notably, one of the five ``BHNS'' alerts is the now
published GW190814 event~\cite{Abbott:2020khf}. In that system, the lower mass object has a mass $M\in [2.51,2.67]M_\odot$ (at 90\% confidence),
and could be either the lowest mass black hole or the highest mass neutron star observed to date. On the other hand, the second likely binary
neutron star system published by the LVC (GW190425)~\cite{Abbott:2020uma} has a most massive object with mass $M\in[1.61,2.52]$ (allowing for non-negligible spins),
and could potentially be a BHNS merger if $(2-3)M_\odot$ black holes exist. It is quite likely that
we will only be sure of the detection of a BHNS binary when we observe either a system where both masses provide unambiguous information
about the nature of the merging objects (a statement that clearly depends on one's priors for the possible distribution of black hole and neutron star
masses), or when we observe an electromagnetic counterpart to a GW event for which the most massive object is guaranteed to be
a black hole (e.g. with $M\gtrsim 3M_\odot$).

Nearly all GW observations performed so far relied on the availability of accurate signal template banks.\footnote{The 
first event, GW150914, was loud enough to be detected with a less model-dependent pipeline~\cite{Abbott:2016blz}} This may lead to some complications and observation
biases when observing BHNS binaries. Indeed, mixed binaries are likely to have larger mass asymmetries and to exhibit more orbital plane and spin precession than black hole binaries
and neutron star binaries. The use of aligned-spin templates in detection pipelines can then lead to the loss of a significant fraction of events~\cite{Harry:2013tca}. GW templates including
precession, tidal effects, and the potential disruption of the neutron star may help alleviate these issues, and will certainly be valuable to perform parameter estimation.
Existing models for the GW signals emitted by BHNS binaries, however, have only recently begun to include both phase and amplitude corrections associated with
tidal distortion and with the disruption of the neutron star (see e.g.~\cite{Lackey:2013axa,Thompson:2020nei,Matas:2020wab}), and do not so far account for precession.

This is where numerical simulations play an important role: analytical models are tested
and calibrated on numerical simulations, to make sure that the models properly capture the late-time inspiral and non-linear merger phase. Numerical waveforms hybridized
with analytical models at early times can also be injected into parameter estimation pipelines to estimate model biases (see e.g.~\cite{Chakravarti:2018uyi,Huang:2020pba} 
for BHNS mergers). There is, however,
a limited number of available numerical waveforms to perform these tests. The $134$ SACRA simulations used by Lackey {\it et al}~\cite{Lackey:2013axa} still provide the
most extensive parameter space coverage of BHNS mergers, yet these waveforms are now quite old. The limited length and accuracy of the simulations make them most
useful to calibrate amplitude corrections at the time of merger, but not as useful to the modeling of the GW phase. Additionally, these waveforms are limited
to aligned BH spins with dimensionless spins $\chi<0.75$. We have recently published a much smaller set of $5$ longer, more accurate SpEC simulations~\cite{Foucart:2018inp},
publicly available as part of the SXS catalogue,\footnote{https://data.black-holes.org/waveforms} but these simulations are limited to non-spinning black holes and obviously do not come close to the
SACRA waveforms in term of parameter space coverage.\footnote{The SACRA code is also capable of generating longer, more accurate waveforms, 
as demonstrated for binary neutron star mergers~\cite{2017PhRvD..96h4060K}} Two of the five have very high neutron star spins, to help efforts to model dynamical tides
in BHNS mergers~\cite{Hinderer:2016eia}, but are otherwise less useful to calibrate models within the most likely range of BHNS parameters. 

In this paper, we present a new set of $3$ BHNS binaries~\cite{SXS_BHNS_0008, SXS_BHNS_0009, SXS_BHNS_0010} performed with the SpEC code. These simulations complement our existing set of long,
accurate simulations. All of these systems have more
realistic mass ratios than most of our public waveforms ($Q=3,4$). Two have higher BH spins than existing public waveforms ($\chi_{\rm BH}=0.9$),
aligned with the orbital angular momentum of the binary. The third has a significant BH spin ($\chi_{\rm BH}=0.75$) misaligned with the orbital angular momentum
by $45^\circ$, leading to significant precession of the orbital plane. All three waveforms are long by the standard of BHNS simulations ($26-33$ cycles).
They also have accuracy comparable to the BHNS waveforms in our current catalogue, despite the use of higher mass ratios and higher black hole spins.
In fact, the numerical accuracy of these waveforms is high enough that we have to more carefully analyze the uncertainty associated with
the extrapolation of the GW signal to null infinity. We thus perform a detailed study of waveform extrapolation errors,
and compare for the first time the results of waveform extrapolation to the waveforms obtained using Cauchy Characteristic Extraction (CCE) methods.

We describe our numerical methods in Sec.~\ref{sec:methods}, the resulting waveforms in Sec.~\ref{sec:GW}, and our numerical accuracy in Sec.~\ref{sec:errors}.
A comparison between waveform extrapolation and CCE is provided in Sec.~\ref{sec:CCE}. In the rest of this paper, we use units such that $G=c=1$, and define
$M_{\rm BH},M_{\rm NS}$ as the ADM masses of the black hole and neutron star at infinite separation, $M=M_{\rm BH}+M_{\rm NS}$ as the total mass of the system,
and $\chi_{\rm BH}$ as the dimensionless black hole spin. All neutron stars in our simulations are initially non-spinning.

\section{Methods}
\label{sec:methods}

\subsection{Evolution methods}

The simulations presented here are performed with the SpEC numerical relativity code.\footnote{http://www.black-holes.org/SpEC.html} SpEC evolves Einstein's equations in the
Generalized Harmonic formalism~\cite{Lindblom:2007} on a pseudospectral grid. The grid rotates and contracts to follow the evolution of the binary, and is distorted so that the apparent
horizon of the black hole remains nearly spherical~\cite{Hemberger:2012jz}. A sphere of constant grid-frame radius is excised from the grid to avoid evolving the interior of the black hole.
The general relativistic equations of hydrodynamics are solved on a separate cartesian grid~\cite{Duez:2008rb,Foucart:2013a}.
Our latest algorithm follows the prescriptions of Radice {\it et al}~\cite{Radice:2012cu} to obtain high-order convergence in smooth regions while capturing shocks. 

The neutron star matter is described by a $\Gamma=2$ ideal gas equation of state, with an ad-hoc thermal component: $P=101.45\rho_0^\Gamma + \Gamma \rho_0 T$, with $P$ the pressure and $\rho_0$ the baryon density.
 We choose the central
density of the neutron star to get a small but reasonable compactness $C_{\rm NS}=GM_{\rm NS}/(R_{\rm NS}c^2)=0.144$ for $M_{\rm NS}=1.4M_\odot$. The use of such a simple equation of state has a few advantages,
including lower simulation costs and higher numerical accuracy than for more realistic models, and the possibility to rescale the result of the simulations with the total mass of the system. Hence, in
this paper, we typically report all masses, times, and distances as dimensionless numbers. Its main disadvantage is that while the dimensionless tidal deformability of the neutron star is 
reasonable ($\Lambda=791$, 
around the upper bound allowed by current observations for low-mass neutron stars), the internal structure of the star and the inferred mass-radius relationship are not. This is less of an issue for BHNS systems than
for binary neutron star systems (as the latter require us to construct two neutron stars, potentially of different masses, with physically consistent tidal properties), but would certainly be a major 
limitation if we wanted to study the formation of a post-merger accretion disk, or any microphysics. For GW modeling, the tidal deformability has been shown
to be the main parameter setting the properties of GW signals both before and during merger~\cite{Lackey:2013axa}. Further tests of the impact of the equations of state beyond the dimensionless
tidal deformability would however be desirable in the future.

We use a third-order Runge-Kutta method for the time evolution. At the end of each time step,
the metric and its derivatives are interpolated from the pseudospectral grid to the finite difference grid, while the fluid variables are interpolated from the finite difference grid to the 
pseudospectral grid. At other times (e.g. at intermediate steps of the Runge-Kutta algorithm, or when evaluating variables using dense output), variables evolved on another grid are
evaluated using linear extrapolation, using their last two communicated values. Overall, the simulations presented here use the exact same numerical methods as the simulations published in
Foucart {\it et al}~\cite{Foucart:2018inp}. 

\subsection{Grid structure}

\begin{figure*}
\includegraphics[width=0.3\textwidth]{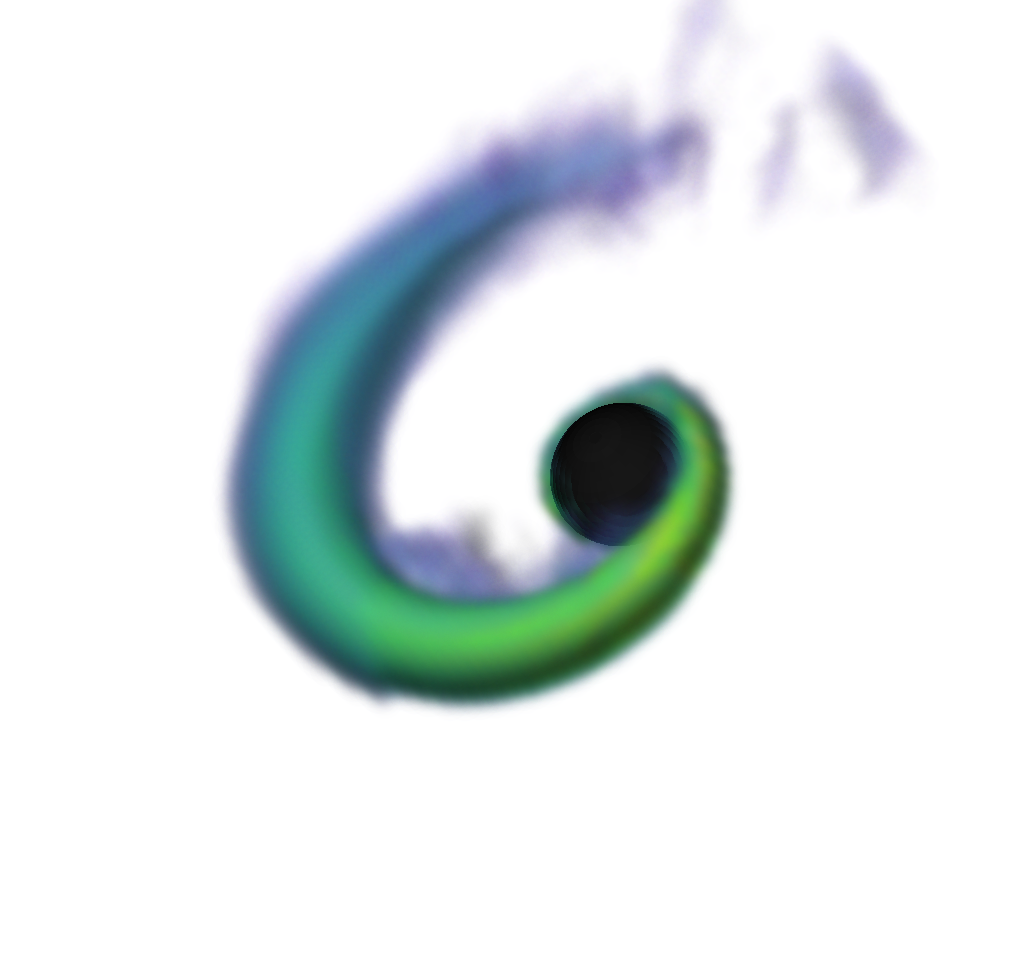}
\includegraphics[width=0.3\textwidth]{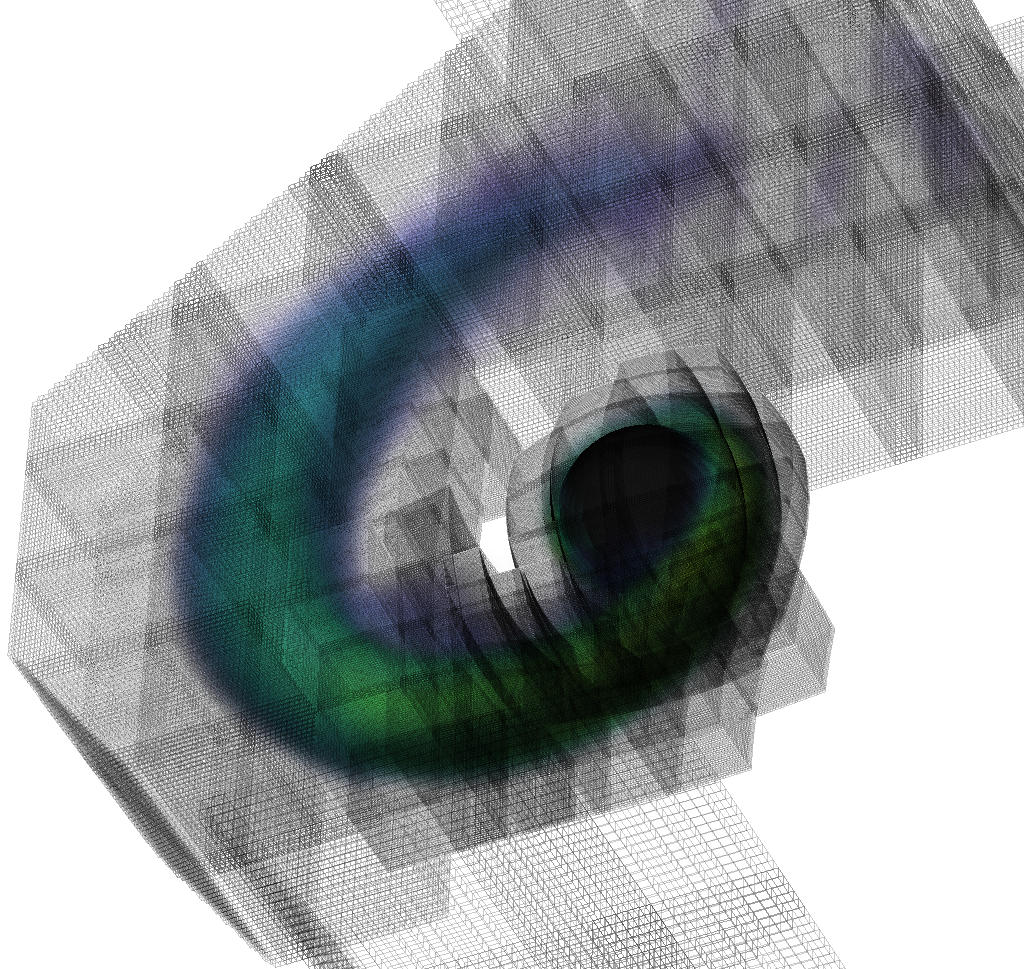}
\includegraphics[width=0.3\textwidth]{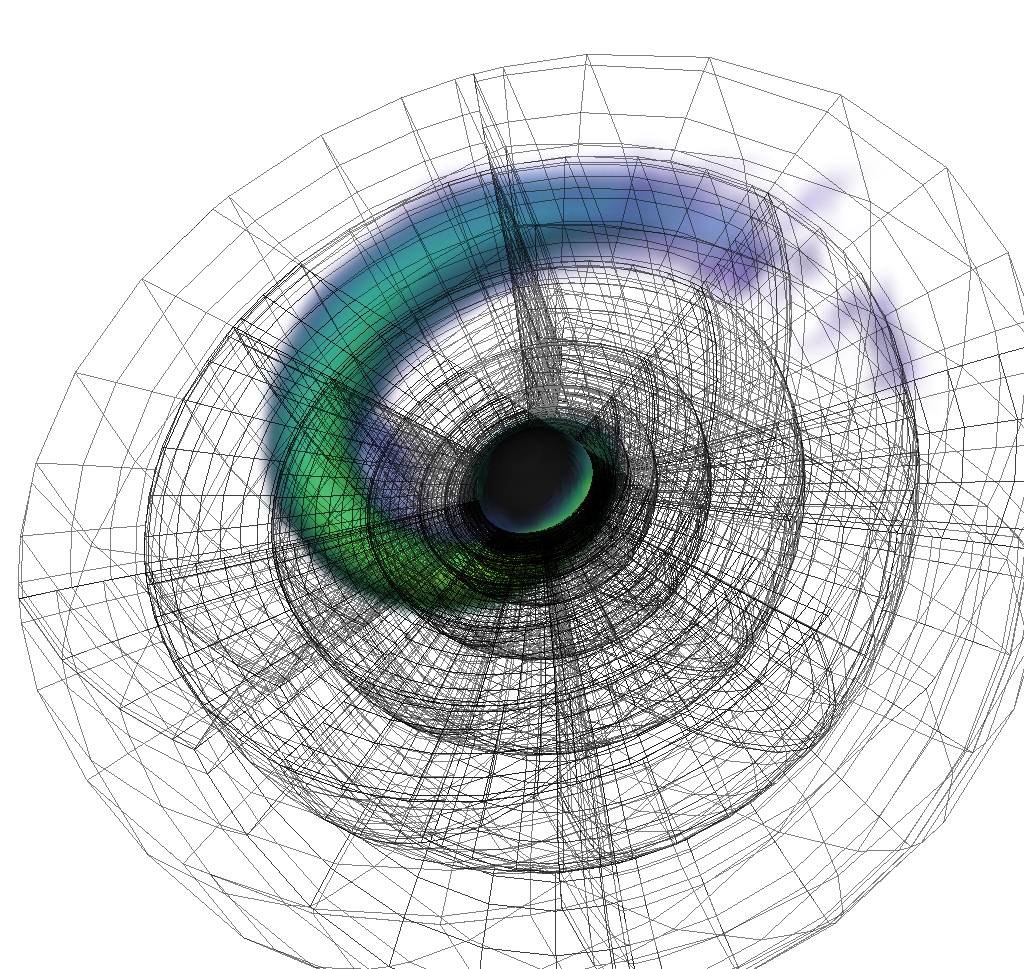}\\
  \caption{Simulation Q4S9 at a time when $\sim 30\%$ of the neutron star mass has been accreted by the black hole. {\it Left}: Matter with density above $6\times 10^7\,{\rm g/cm^3}$, and apparent horizon of the black hole. {\it Center}: Finite difference grid,
  showing both mesh refinement and the fact that the grid does not cover vacuum regions. {\it Right}: Inner region of the pseudospectral grid, showing mesh refinement close to the black hole and in regions where dense matter is present. We only show points below the orbital plane and within $\sim 15M$ of the black hole's center. The pseudospectral grid extends to much larger distances ($500M$), using spherical shells not shown on the figure. All three figures are taken from our lowest resolution simulation.}
  \label{fig:grids}
\end{figure*}

The finite difference grid used in our simulations has a relatively simple structure. Before merger, we use a Cartesian grid with constant grid spacing $\Delta x_{\rm grid}$ in the grid coordinates.
As the grid contracts during the binary inspiral, this would lead to a significant decrease in the grid spacing in the inertial frame, to $0.4\Delta x_{\rm grid}$ by the time of merger,
dramatically increasing the cost of the simulations. Instead, whenever the grid spacing decreases by $20\%$ in the inertial frame, we reset it to the grid spacing at the initial time, interpolating onto 
a coarser finite difference grid. This operation is performed $\sim 4-5$ times per simulation, and keeps the resolution roughly constant during the evolution. Once the neutron star disrupts, we use
fixed mesh refinement. The finest level of refinement is a grid of $324^3$ cells, centered on the black hole, and with the same spacing as the pre-merger grid in the inertial frame. Additional levels of
refinement are added as needed, each new level a $324^3$ grid centered on the black hole and twice the grid spacing of the previous level. We note that to save computational resources, these
grids are divided into $27^3$ cells blocks that can be fully ignored by the evolution if no matter is present in the region that they cover. After the gravitational wave signal from the merger
has left the finite difference grid, we save computational resources by reducing the resolution of that grid: by that point, following the gravitational waves to the radius at which they are extracted is our
only concern, and this only requires evolution of Einstein's equations. We evolve each configuration at $3$ resolutions, summarized in Table~\ref{tab:ID}. If we assume $M_{\rm NS}=1.4M_\odot$,
these correspond to initial grid spacings $\Delta x_{\rm FD}^0=(295,236,189)\,{\rm m}$.

Before merger, the pseudospectral grid is constructed from 8 spherical shells surrounding the black hole, 1 ball and 8 spherical shells covering the neutron star and its surroundings, and 32 spherical shells covering the wave region
(centered on the center of mass of the binary, and with radii ranging from 2.5 times the binary separation to $R_{\rm out}=500M$). Between these 3 regions, we use distorted cylinders, with the line connecting the compact objects as their axis. After merger, the region inside of the 32 outer shells is covered with ``CubedSphere'' subdomains, i.e. cubes distorted so that one
coordinate is constant at constant radius (defined as the distance to the center of the remnant black hole). The number of basis functions within each of these subdomains is chosen adaptively, to obtain a user-specified
accuracy estimated from the spectral coefficients of the evolved variables~\cite{Szilagyi:2014fna,Foucart:2013a}. The target accuracy on the pseudospectral grid scales as $(\Delta x_{\rm FD}^0)^5$. Errors on the pseudospectral grid should thus converge to zero faster than the errors on the finite difference grid.

Visualizations of the matter density, black hole apparent horizon, finite difference grid, and pseudospectral grid are provided in Fig.~\ref{fig:grids} for simulation Q4S9 (see next section), around the time of merger. The figure illustrates the adaptivity of both numerical grids.

\subsection{Initial Conditions}

Initial conditions for all simulations are obtained using our in-house Spells initial data solver~\cite{Thesis:Pfeiffer,Pfeiffer2003,FoucartEtAl:2008}. Spells solves for the constraints in Einstein's equations and for an irrotational velocity profile inside
the neutron star, while imposing hydrostatic equilibrium. We first find initial data on a quasi-circular trajectory~\cite{Pfeiffer2000,FoucartEtAl:2008}, then reduce the eccentricity according to the iterative procedure described
in Pfeiffer {\it et al}~\cite{Pfeiffer-Brown-etal:2007}. Each iteration requires the evolution of the binary for $\sim 3$ orbits. We target eccentricities of $\lesssim 0.002$ for non-precessing systems. For the precessing system in this paper, the eccentricity of the
quasi-circular initial data was already very small ($e\sim 0.003$), and could not be reduced using our standard procedure.

\begin{figure}
\includegraphics[width=0.95\columnwidth]{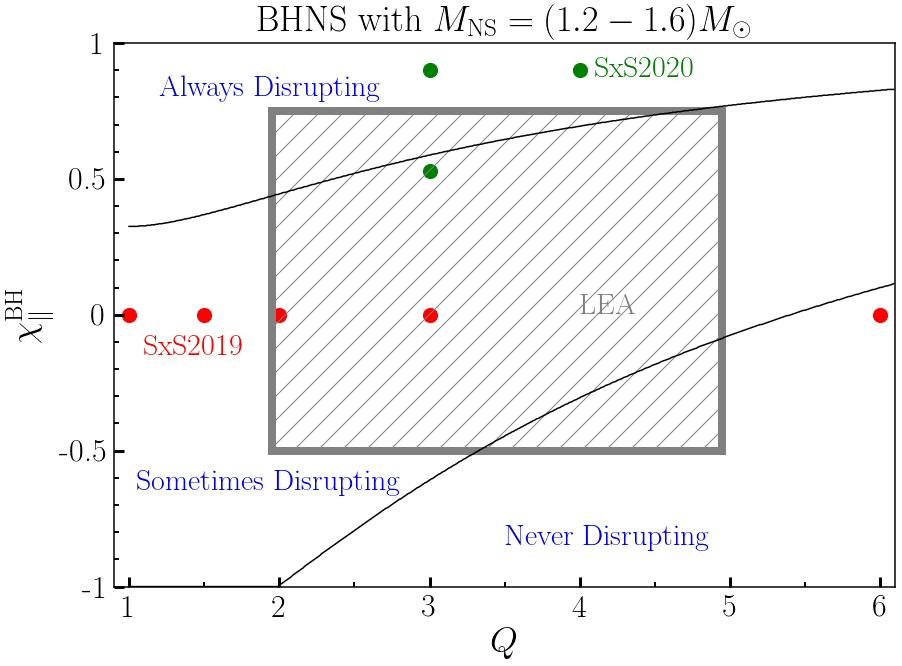}\\
  \caption{Distribution of the BHNS simulations in the SXS catalogue projected in the $Q$-$\chi_\parallel^{\rm BH}$ plane. Simulations from~\cite{Foucart:2018inp} are red dots (SXS2019), while simulations from this work are green dots (SXS2020). The region
  of parameter space covered by 134 short waveforms from~\cite{Lackey:2013axa} is shown in grey (LEA). For context, we also
  show the regions of parameter space where $(1.2-1.6)M_\odot$ neutron stars satisfying the equations of state constraints of~\cite{GW170817-NSRadius} always disrupt, disrupt for some equations of state and/or neutron star masses only, or never disrupt,
  according to the fitting formula from~\cite{Foucart:2018rjc}.}
  \label{fig:Overview}
\end{figure}

\begin{table}
\caption{
  Overview of the simulations presented in this paper.
  $M_{\rm BH,NS}$ are the ADM masses of the BH and NS in isolation, 
  $\chi_{\rm BH}$ the initial dimensionless BH spin, $e$ the initial eccentricity, $i_{\rm BH}$ the
  initial inclination of the BH spin with respect to the orbital angular momentum vector,
  $\tilde \Lambda$ the effective dimensionless tidal deformability of the binary, $\Omega_0$ the initial
  angular velocity, $M=M_{\rm BH}+M_{\rm NS}$ the total mass, $t_{\rm peak}$ the
  time at which the $(2,2)$ mode of the GW signal reaches its maximum amplitude, and 
  $\Delta x_{\rm FD}^{t=0}$ the initial spacing of the finite volume grid.
}
\label{tab:ID}
\begin{tabular}{c|c|c|c|c|c|c|c|c}
{\rm Name} & $\frac{M_{\rm BH}}{M_{\rm NS}}$  & $\chi_{\rm BH}$ & $e$ &  $i_{\rm BH}$ & $\tilde \Lambda$ & $\Omega_0 M$ & $\frac{t_{\rm peak}}{M}$ & $\frac{\Delta x_{\rm FD}^{t=0}}{M_{\rm NS}}$\\
\hline
Q3S9-L0 &  3 & 0.9 &0.0004 & $0^\circ$ & 35.2 & 0.0236 & 2342.6 & 0.143\\
Q3S9-L1 &  3 & 0.9 &0.0005 & $0^\circ$ & 35.2 & 0.0236 & 2346.8 & 0.114\\
Q3S9-L2 &  3 & 0.9 &0.0005 & $0^\circ$ & 35.2 & 0.0236 & 2346.5 & 0.091\\
\hline
Q4S9-L0 &  4 & 0.9 &0.0017 & $0^\circ$ & 15.3 & 0.0243 & 2661.7 & 0.143\\
Q4S9-L1 &  4 & 0.9 &0.0018 & $0^\circ$ & 15.3 & 0.0243 & 2660.1 & 0.114\\
Q4S9-L2 &  4 & 0.9 &0.0017 & $0^\circ$ & 15.3 & 0.0243 & 2660.2 & 0.091\\
\hline
Q3S75p-L0 &  3 & 0.75 &0.0031 & $45^\circ$ & 35.2 & 0.0197 & 3473.5 & 0.143\\
Q3S75p-L1 &  3 & 0.75 &0.0031 & $45^\circ$ & 35.2 & 0.0197 & 3469.9 & 0.114\\
Q3S75p-L2 &  3 & 0.75 &0.0031 & $45^\circ$ & 35.2 & 0.0197 & 3472.3 & 0.091\\
\end{tabular}
\end{table}

We consider 3 initial configurations, summarized in Table~\ref{tab:ID}.Figure~\ref{fig:Overview} also provides an overview of these simulations and of existing public BHNS waveforms. The first configuration is a system with mass ratio $Q=3$ and aligned BH spin $\chi_{\rm BH}=0.9$, hereafter named Q3S9. The second is identical except for the choice of a mass ratio $Q=4$, and is named Q4S9. The last configuration has a dimensionless BH spin $\chi_{\rm BH}=0.75$, misaligned by $45^\circ$ with the orbital angular momentum, and initially in the plane formed by the orbital angular momentum vector and the line connecting the center of the compact objects. The misalignment of the spin leads to significant precession of the orbital plane of the binary. We label this simulation Q3S75p.
The initial orbital frequencies are chosen to provide more than $\sim 12$ orbits before merger, and we find indeed that simulation Q3S9 evolves for $13.2$ orbits before the peak of the dominant $(2,2)$ mode of the GW signal, Q4S9 for $15.6$ orbits, and Q3S75p for $16.3$ orbits. By this metric, these 3 simulations are longer than all but one of the existing public BHNS waveforms.\footnote{The longest public BHNS waveform is a non-spinning, $Q=1.5$ simulation evolved for $16.6$ orbits}~\cite{Foucart:2018inp} In terms of the number of time steps required, which may be more relevant to the growth of numerical errors, the simulations presented here are significantly longer than any public BHNS waveform.

As shown on Fig.~\ref{fig:Overview}, the initial conditions for our simulations are all in the regime where the neutron star is strongly disrupted by the black hole -- and would be disrupted even for softer equations of state. The two high-spin simulations
are also out of the range of the SACRA simulations used by Lackey {\it et al} (LEA)~\cite{Lackey:2013axa}, which makes them useful to test models of both the phase evolution and tidal disruption of BHNS mergers. Our lower spin waveform, on the other hand, has the advantage of being the only precessing system shown on Fig.~\ref{fig:Overview}. Less effort has gone into the production of waveforms for non-disrupting binaries, in part because our best simulation in that regime~\cite{Foucart:2013psa}, the $Q=6$ simulation on Fig.~\ref{fig:Overview}, showed that the resulting waveform could not be distinguished from an equivalent BBH waveform, at least within our numerical errors (or with any existing GW detector).

\subsection{Waveform extraction}

We use two independent methods to estimate the gravitational wave signal at infinity from the values of the metric at finite radii. The first, used in all of our previous BHNS publications, follows the procedure
outlined by Boyle \& Mroue~\cite{Boyle-Mroue:2008}. The Newman-Penrose scalar $\Psi_4$ and metric perturbation $h$ are estimated on spheres of constant inertial radii $R_i$ (the latter using Regge-Wheeler-Zerilli
techniques), and decomposed into spin=$-2$ spherical harmonics components. For each $R_i$, we then compute a retarded time $t_{\rm ret}(t,R_i)$ approximately accounting for the travel time of the wave
from the merging compact objects to $R_i$. We then fit the ansatz 
\beqn
A_{lm}(t_{\rm ret},r)&=&\sum_{j=0}^N A_{lm,j}(t_{\rm ret})r^{-j}\\
\phi_{lm}(t_{\rm ret},r)&=&\sum_{j=0}^N \phi_{lm,j}(t_{\rm ret})r^{-j}
\eeqn
to the amplitude $A_{lm}$ and phase $\phi_{lm}$ of the $(l,m)$ component of our spherical harmonics decomposition, at a fixed set of retarded times. The $(l,m)$ mode at infinity is then estimated to be $A_{lm,0}e^{i\phi_{lm,0}}$.
This procedure can be applied to either $\Psi_4$ or $h$. In our simulations, we fit this ansatz to the estimated values of $h,\Psi_4$ at $20$ radii between $100M$ and $400M$, equally spaced in $r^{-1}$.

The second method is used here for the first time in our fluid simulations. We perform Cauchy Characteristic Evolution (CCE)~\cite{Bishop1996,Bishop:1997ik,Winicour2005,Babiuc:2010ze,Handmer:2014} using the methods described in Moxon {\it et al}~\cite{Moxon:2020gha}, and implemented in the open-source SpECTRE code.\footnote{https://github.com/sxs-collaboration/spectre} In CCE, we use the evolution data on a surface of constant inertial radius ($R=200M,300M,400M$ here) as boundary condition for a non-linear evolution on a null foliation of the spacetime outside of that surface, propagating the signal to null infinity. CCE carries to null infinity the Bondi news, and the SpECTRE implementation of CCE also provides estimates of all five Weyl scalars and of the gravitational wave strain -- although with some caveats related to initial data and Bondi-Metzner-Sachs (BMS) freedom for the latter, discussed in Sec.~\ref{sec:CCE}.

With these two methods at our disposal, we can more carefully study the reliability of our estimates of the signal at null infinity.

The waveforms publicly released as part of the SXS catalogue include two versions of the extrapolated signal. The first contains the waveform in the inertial frame of the simulation. The second corrects for the motion of the center-of-mass of the binary, and provides the waveform in the rest frame of the system (before merger)~\cite{Woodford:2019tlo,SXSCatalog2018}. This procedure avoids some mode mixing in all cases, and is particularly useful for the precessing system: precessing BHNS binaries generated with our initial data solver have non-zero velocity in the direction perpendicular to the initial orbital plane ($\sim 0.001c$) that can lead to significant phase errors at merger ($\sim 0.3\,{\rm rad}$) simply due to the change in the time-of-flight of the waveforms from the binary to the observer. Comparisons between numerical results and analytical models presented in this paper are performed after removal of the center-of-mass motion.

\section{Results}

\subsection{Overview}
\label{sec:GW}

\begin{figure*}
\includegraphics[width=0.95\textwidth]{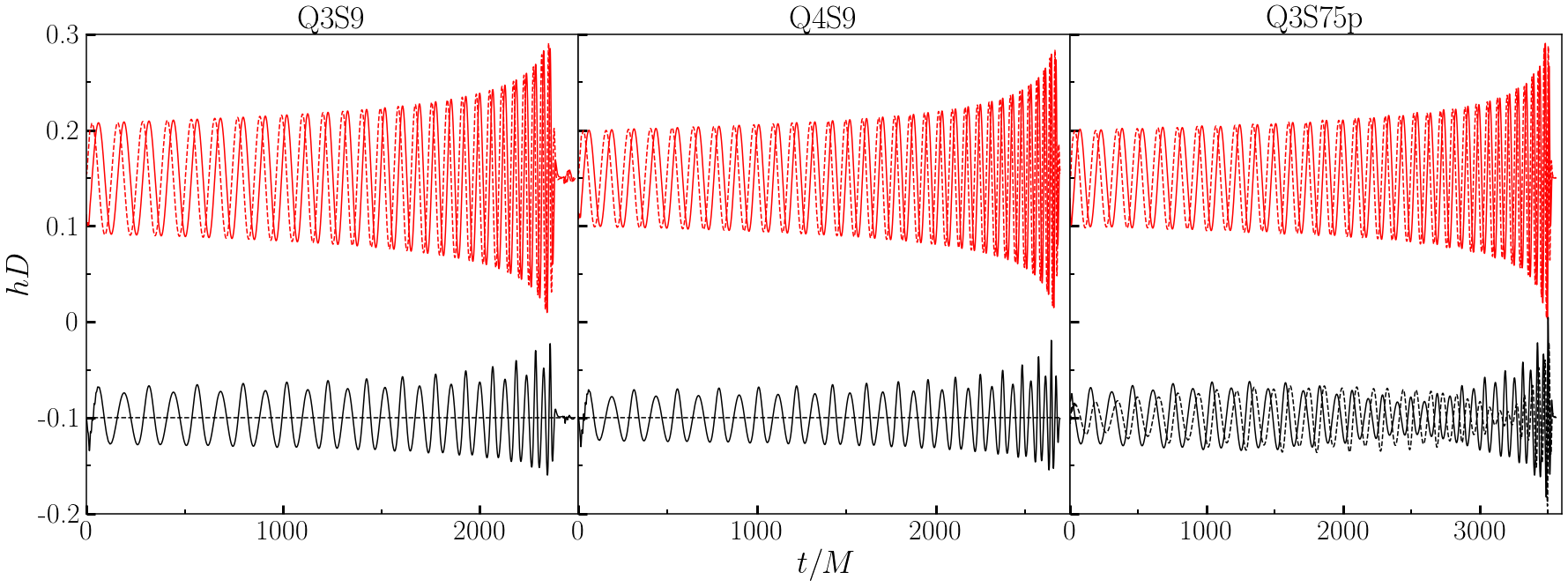}\\
  \caption{GW strain for simulations Q3S9 (left), Q4S9 (center) and Q3S75p (right). Each plot shows results for an observer in the direction of the total angular momentum (top, `face-on' for non-precessing
  system) and from a direction that, at $t=0$, is orthogonal to the total angular momentum and within the orbital plane of the binary (bottom, `edge-on' at $t=0$).
  We show both polarizations $h_+$ (solid lines) and $h_\times$ (dashed lines), and shift the average value of the strain for readability. The edge-on signals clearly show the impact of mass
  asymmetry (oscillations in the maximum of the strain, due to a significant $l=3,m=3$ mode) and, for Q3S75p, precession (slow oscillation in the amplitude of the signal, non-zero value of $h_\times$). Q3S75p
  goes through slightly more than half a precession cycle over the course of the simulation.}
  \label{fig:WavesInDir}
\end{figure*}

The evolution of all three configurations proceed as is typical for BHNS systems with moderate mass ratios $Q\sim 2-4$ and significant spins: the neutron star disrupts well out of the innermost stable circular
orbit of the black hole, leading to mass ejection and the formation of a massive accretion disk (see Figs.~\ref{fig:grids}-\ref{fig:Overview}). In these simulations, however, we do not attempt to follow the evolution of the post-merger
remnant with enough accuracy to properly measure the masses of the ejecta and post-merger accretion disks, in part due to the high computational cost of following rapid accretion by the black hole in BHNS systems, and in part because
the post-merger evolution likely lacks realism when using a simple $\Gamma=2$ ideal gas equation of state. Once the neutron star disrupts, stopping GW emission, we focus on following the propagation of the GW signal to large distances.

The extrapolated GW signals obtained from our simulations are publicly available as part of the SXS catalogue, for extrapolation orders $N=2-5$ and for multipoles up to $l=8$.
Figure~\ref{fig:WavesInDir} shows the $h_+$ and $h_\times$ polarizations of the GW signals, for observers who are, at $t=0$, in the direction of the total angular momentum of the binary, or perpendicular to that direction and
in the orbital plane. The clearest
difference between these BHNS waveforms and equivalent BBH waveforms is the rapid cutoff in GW emission when the neutron star disrupts. The other
features of the signal are similar to BBH systems: the signal in the direction of the total angular momentum of the system
is entirely dominated by the $l=2,m=\pm 2$ modes and is thus a nearly feature-less chirping signal. For non-precessing systems, the edge-on signal
shows more clearly the impact of unequal masses. For Q3S75p, in addition to the mass asymmetry, the precession of the orbital plane is clearly visible for the initially `edge-on' observer. 
The $h_\times$ signal, in particular, clearly shows that the Q3S75p system goes through slightly more than half a precession cycle during the simulation.

An important aspect of the BHNS binary systems presented here, as opposed to those previously published in the SXS catalogue~\cite{Foucart:2018inp}, is the impact of subdominant modes on the signal.
For the Q3S9 (resp. Q4S9) configuration, the peak amplitude of the $(3,3)$ mode of the strain is $18\%$ (resp. $21\%$) of the peak amplitude of the dominant $(2,2)$ mode. The $(4,4)$ mode
has $7\%$ ($8\%$) of the peak amplitude of the dominant mode, while other modes remain below $5\%$ of the amplitude of the $(2,2)$ mode. For Q3S75p, the precession of the orbital plane additionally leads to
the mixing of modes with the same $l$ but different $m$. All $l=2$ and $l=3$ modes then have significant amplitude. Accordingly, these new public waveforms 
should be particularly useful to test the effect of higher-order modes on detection and parameter estimation for BHNS binaries.

\subsection{Numerical accuracy}
\label{sec:errors}

\begin{figure}
\includegraphics[width=0.95\columnwidth]{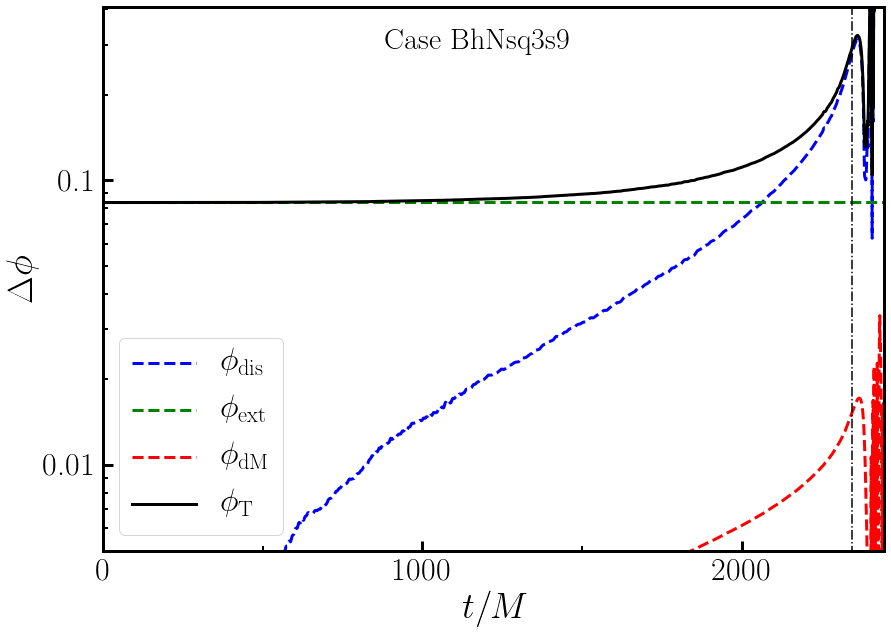}\\
\includegraphics[width=0.95\columnwidth]{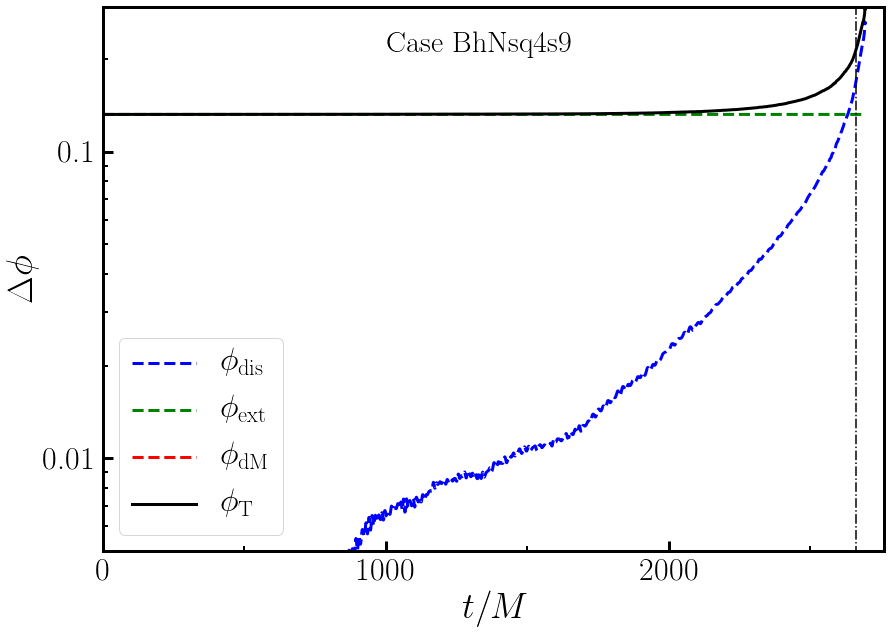}\\
\includegraphics[width=0.95\columnwidth]{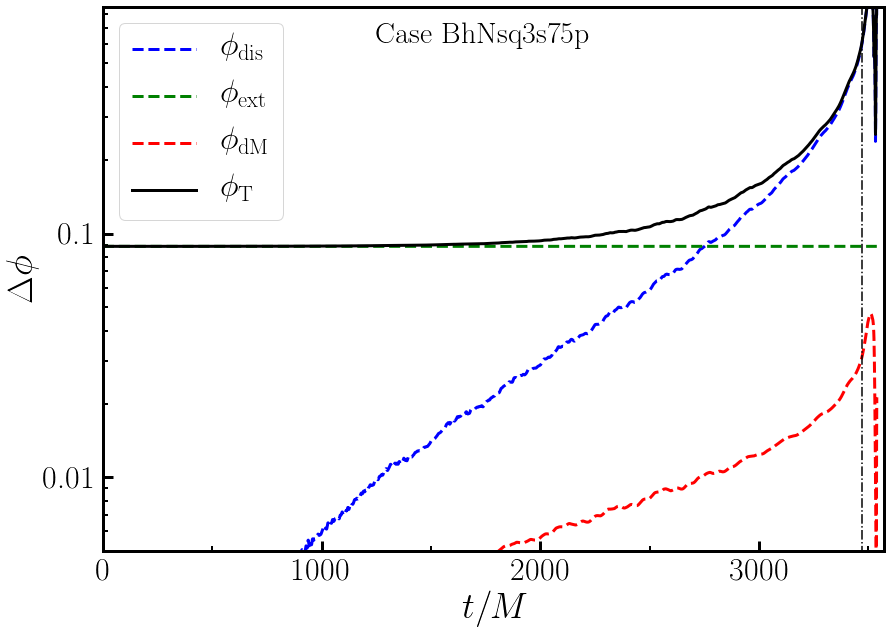}
  \caption{Estimates of the phase error for the $(2,2)$ mode of the GW signal in each simulation. We include the impact of mass losses ($\Phi_{\rm dM}$),
  extrapolation error ($\Phi_{\rm ext}$), and grid resolution ($\Phi_{dis}$), following the methods described
  in~\cite{Foucart:2018inp} and Sec.~\ref{sec:errors}. The vertical dot-dashed curves indicate $t_{\rm peak}$.}
  \label{fig:StErrPhi}
\end{figure}

The main expected uses of our waveforms are testing and calibrating analytical waveform models, and injection in detection and parameter estimation pipelines used by GW observers.
It is thus critical to provide careful, conservative estimates of our errors, to avoid introducing systematic biases in these studies, and in upcoming observations of BHNS binaries. In
Foucart {\it et al}~\cite{Foucart:2018inp}, we proposed a standardized method to estimate phase errors in SpEC BHNS waveforms that accounts for $3$ main sources of error: finite
resolution, extrapolation of the signal to infinity, and mass loss at the boundary of the finite difference grid used to evolve neutron stars. We summarize this method here, and plot the
resulting error estimates on Fig.~\ref{fig:StErrPhi}, for the dominant $(2,2)$ mode of the signal.

Finite-resolution errors are estimated by comparing the results of our low (L0), medium (L1), and high (L2) accuracy simulations. We first measure the phase differences between L0 and L2, 
and estimate the error in the L2 waveform using Richardson extrapolation to infinite resolution. The extrapolation is performed assuming 2nd-order convergence, a relatively conservative estimate
considering that the methods used in our simulations are typically at least 3rd order convergent. We then perform the same calculation, but using the results of the L1 and L2 simulations. At any
time, the worst of these two error estimates is assumed to be the finite-resolution error. We note that we need this comparison because in the hybrid spectral/finite volume algorithm used in SpEC, 
different parts of the evolution may dominate the error budget at different times. Unfortunately, as a result the sign of the phase difference between two simulations may change over the course
of the evolution, leading to occasional cancellations of the error estimates based on 2 resolutions only. On the other hand, we have found that the more complex error estimate described here has provided us with 
conservative estimates of the numerical error whenever improved numerical methods / increased computational resources have allowed us to test it against higher accuracy results.

Extrapolation errors are estimated by measuring the phase difference between 2nd order and 3rd order extrapolation between $t=0$ and $t_{\rm peak}$ (the time when the amplitude of the $(2,2)$ mode of the GW signal
is maximum, see Table~\ref{tab:ID}). The maximum phase difference over that time span is taken as the extrapolation error. This is a very conservative choice that was made largely because, as opposed to BBH simulations,
BHNS simulations do not show clear convergence of the extrapolated waveform with the chosen extrapolation order. This has not been much of an issue so far, as extrapolation errors remained much smaller
than finite resolution errors~\cite{Foucart:2018inp}. Fig.~\ref{fig:StErrPhi} shows that this is no longer the case in these new simulations. We thus perform a more in-depth study of extrapolation errors in the following
section, that indicates that extrapolation using a 2nd order polynomial in $1/r$ leads to extrapolation errors that are significantly smaller than those shown in Fig.~\ref{fig:StErrPhi}.

Finally, a small mass loss at the boundary of our finite difference grid could lead to an error in the mass of the neutron star, and thus in the phase evolution of the system. However, none of the simulations presented here
loses more than $10^{-4}M_\odot$ over the course of the binary inspiral, and the phase error due to mass loss at grid boundaries is thus negligible. 

Overall, we note that
the phase error at $t_{\rm peak}$ is $(0.2-0.4)\,{\rm rad}$ and limited by the finite resolution of the simulation, while during inspiral it is $\lesssim 0.1\,{\rm rad}$ and limited by the estimated extrapolation error.
We will however see in Sec.~\ref{sec:CCE} that the true extrapolation error is nearly certainly significantly smaller than what is shown on Fig.~\ref{fig:StErrPhi}. We keep the estimate on Fig.~\ref{fig:StErrPhi} to
allow for direct comparisons with our previous waveforms~\cite{Foucart:2018inp}. Only one simulation from~\cite{Foucart:2018inp} has smaller phase errors, and it is an equal mass, non-spinning configuration that
is significantly easier for our code to evolve, and slightly shorter (in number of orbits) than the simulations presented here.

\begin{figure}
\includegraphics[width=0.95\columnwidth]{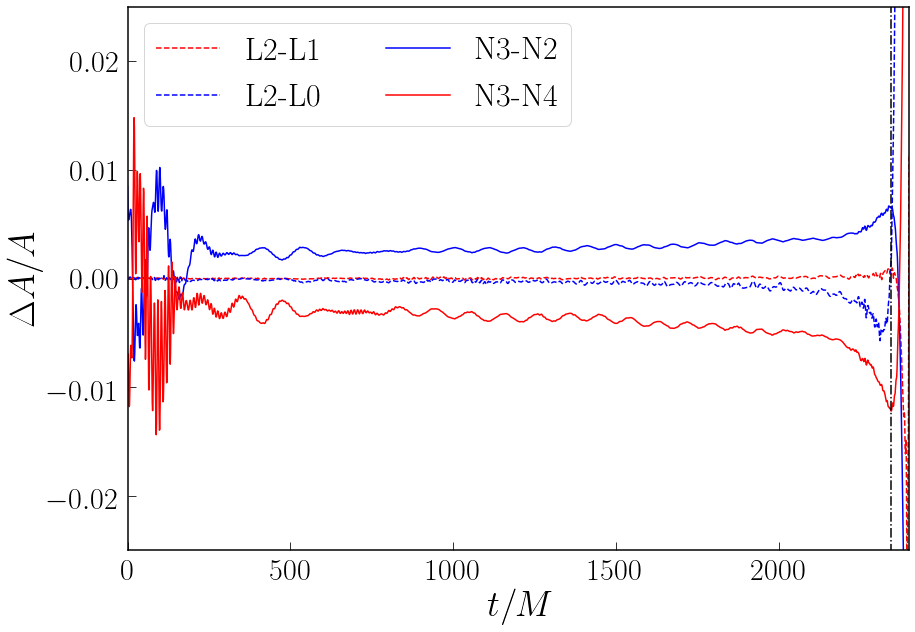}\\
\includegraphics[width=0.95\columnwidth]{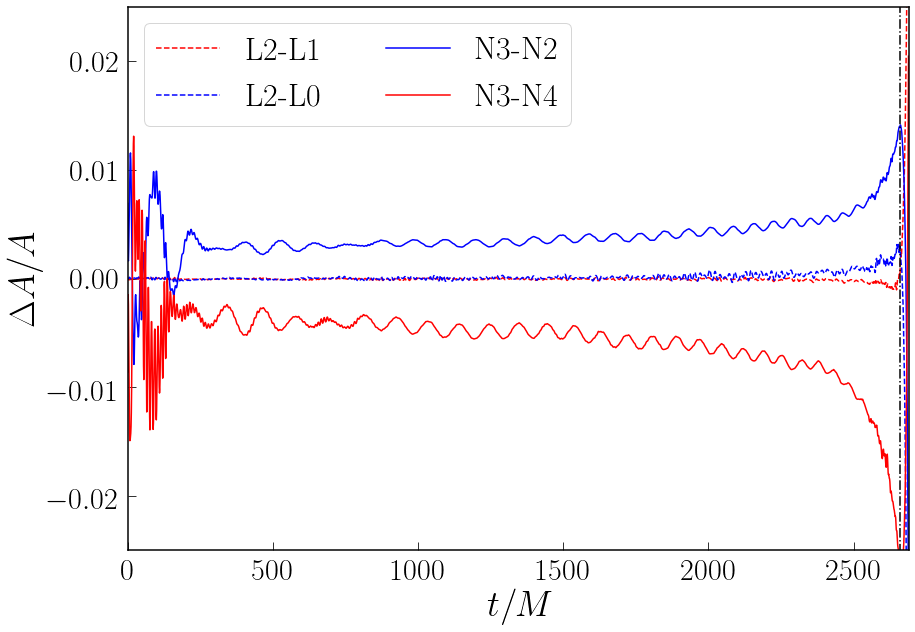}\\
\includegraphics[width=0.95\columnwidth]{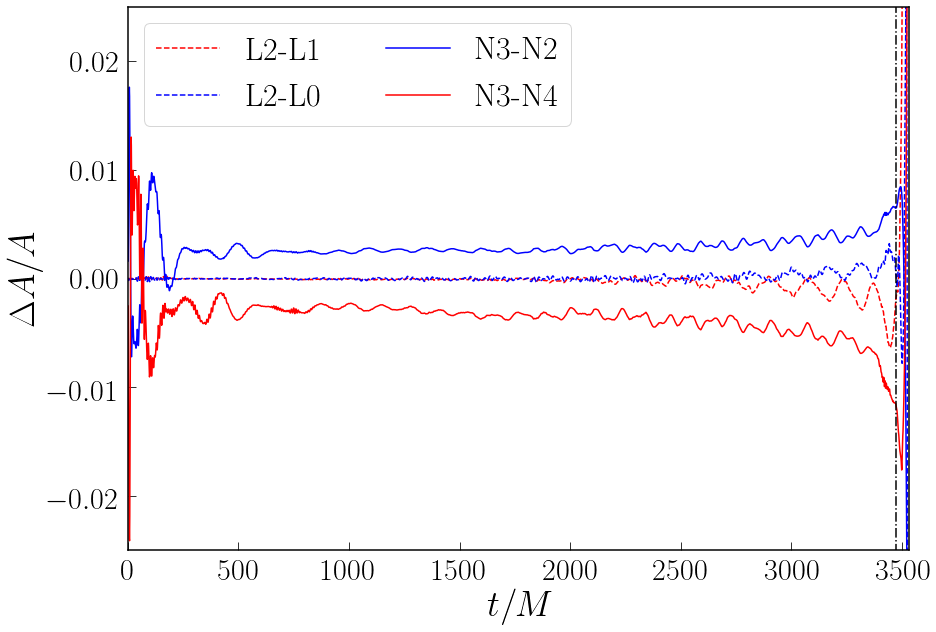}
  \caption{Estimates of the relative error in the amplitude of the (2,2) mode of the GW signal for each simulation. From top to bottom, we show results
  for Q3S9, Q4S9, and Q3S75p. Dashed curves show comparisons between the amplitude obtained with different numerical
  resolutions ($L0,L1,L2$), while the solid curves show comparisons between the amplitude for different extrapolation orders ($N2,N3,N4$).
  The vertical dot-dashed curves indicate $t_{\rm peak}$.
  We clearly see that extrapolation is the main source of error when estimating the GW amplitude. Amplitude errors are $\lesssim 1\%$, except
  for simulation Q4S9 at the time of merger.}
  \label{fig:StErrA}
\end{figure}

We can also estimate the uncertainty in the amplitude of the GW signal in our simulations. Fig.~\ref{fig:StErrA} shows relative differences between the amplitude of the $(2,2)$ mode
of the waveforms at different resolutions, and using different extrapolation orders. We see that the errors are small (typically $\lesssim 1\%$), especially when compared to current 
calibration uncertainties in GW detectors (e.g. $7\%-10\%$ for GW170817~\cite{2017ApJ...848L..13A}). Finite-resolution errors often appear negligible when compared to extrapolation errors.
This is because the main error due to finite resolution is a small shift of the time required for binaries to orbit / inspiral. This has a much larger effect on the phase of the gravitational wave signal
than on its slowly-varying amplitude.

Very similar results are found for higher-order $(l,m)$ modes, except that the phase error is multiplied by $l/2$, i.e. the ratio of the frequency of the $(l,m)$ mode and the frequency of the $(2,2)$ mode. This
is once more a consequence of the fact that the dominant source of error is a slight change in the evolution timescale of the system. The resulting time shift in the waveform is the same for all modes, and the
associated phase error is thus proportional to the frequency of the mode. Relative errors in the amplitude of the signal only increase slightly for higher order modes (see e.g. Figs~\ref{fig:CceVsExtH}-\ref{fig:CceVsExtH-33} 
in the next section). The absolute error in the amplitude of the signal is thus dominated by the error in the dominant $l=2$ mode(s) for most binary orientations.

\subsection{Extrapolation errors and Cauchy Characteristic Evolution}
\label{sec:CCE}

\begin{figure}
\includegraphics[width=0.95\columnwidth]{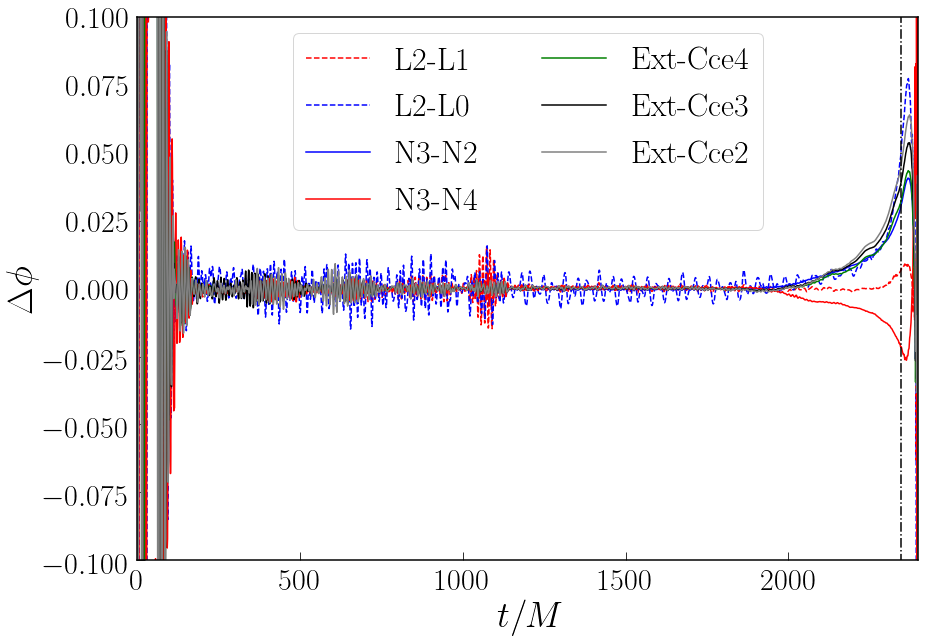}\\
\includegraphics[width=0.95\columnwidth]{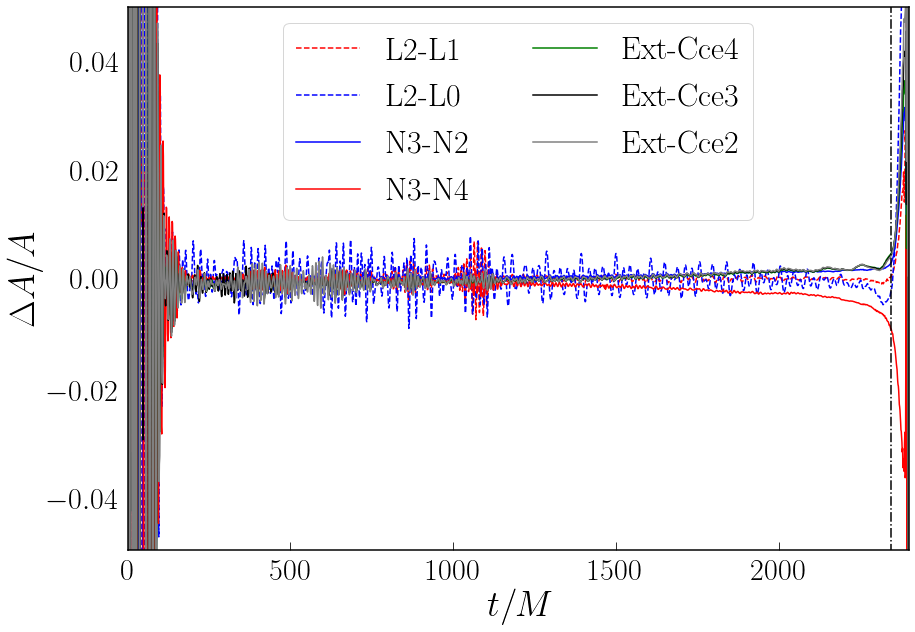}
  \caption{Difference in the phase of the (2,2) mode for various estimates of $\Psi_4$ in system Q3S9, after application
  of a time and phase shift minimizing the phase error in the window $t\in[500,2300]$. 
  We show phase differences between our numerical resolutions (L0,L1,L2),
  between different extrapolation orders (N2,N3,N4, for the highest resolution simulation), and between the highest resolution simulation with $N=3$ and the waveforms
  obtained using CCE from data extracted at $R=200M$ (Cce2), $R=300M$ (Cce3), and $R=400M$ (Cce4). We see very good agreement between the various CCE
  waveforms, and between $N=2$ extrapolation (solid blue curve) and CCE.}
    \label{fig:CceVsExtPsi4Q3}
\end{figure}

\begin{figure}
\includegraphics[width=0.95\columnwidth]{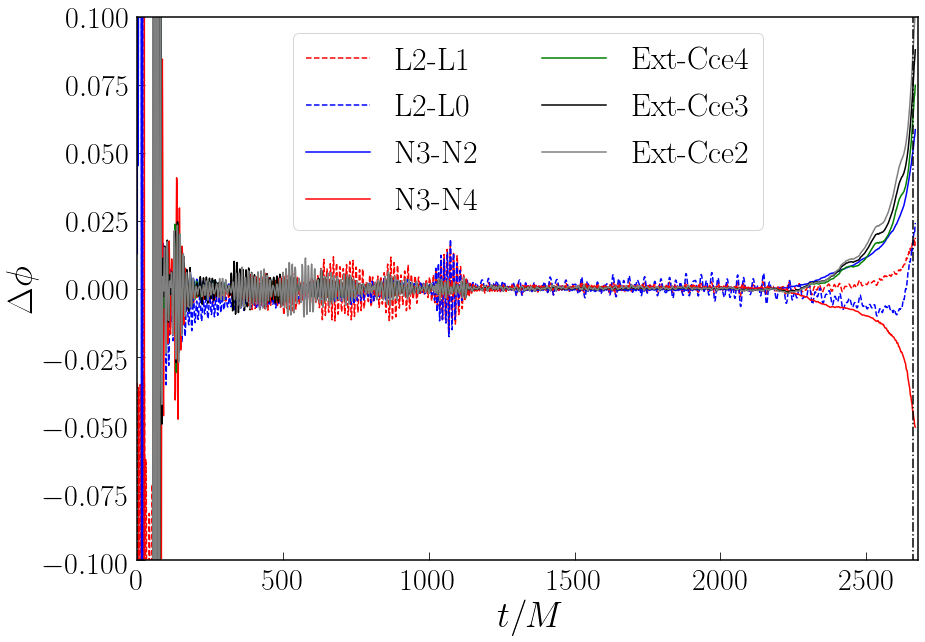}\\
\includegraphics[width=0.95\columnwidth]{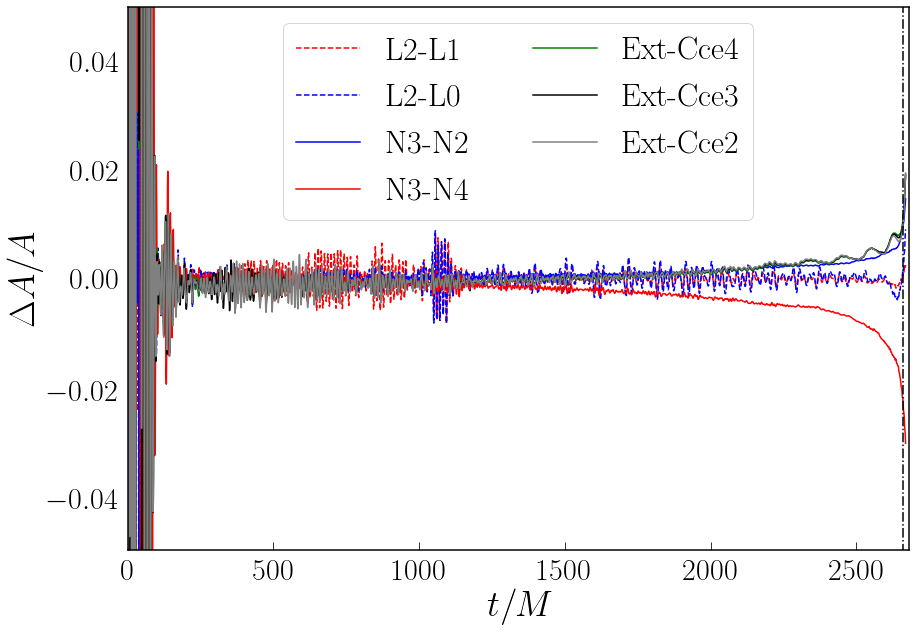}
  \caption{Same as Fig.~\ref{fig:CceVsExtPsi4Q3}, but for the $(2,2)$ mode in the Q4S9 configuration. We find very similar results, including very good agreement between all CCE results and the $N=2$ extrapolation results.}
  \label{fig:CceVsExtPsi4Q4}
\end{figure}

The previous section showed that extrapolation errors may significantly contribute to our error budget. However, extrapolation errors are difficult to assess: there is no clear improvement as the order of extrapolation increases,
and in fact errors tend to grow beyond $N=4$ extrapolation. In this section, we will argue that $N=2$ extrapolation is more accurate than our previous estimates would indicate. The main argument to that effect is a direct
comparison of the GW signals obtained using extrapolation, and those obtained using CCE. We begin with a study of the Weyl scalar $\Psi_4$ (proportional to the second time derivative of the strain), for reasons that will
become clear below. Fig.~\ref{fig:CceVsExtPsi4Q3} shows the phase and amplitude differences between all wave extraction methods for case Q3S9, and Fig.~\ref{fig:CceVsExtPsi4Q4} provides the same information for 
case Q4S9. Differences between simulations at 3 resolutions are also shown for reference. We note that CCE does not provide an absolute value of the time that can be used consistently for all waveforms, due to the unknown
travel time between the inner boundaries of the various CCE evolution systems, and the differences between the simulation time of extrapolation and the asymptotically inertial time of CCE. Accordingly, all comparisons in this section are performed after application of a time and phase shift to the waveforms, chosen to minimize the phase difference
before merger. Phase differences between waveforms using different resolution are naturally smaller than without alignment, to the point that after alignment differences between extrapolation methods are at least of the same magnitude
as differences between numerical resolutions. However, both case show two important results: all CCE waveforms are in very good agreement with each other
($\lesssim 0.01\,{\rm rad}$ phase difference and $\lesssim 0.1\%$ amplitude error), and all CCE waveforms agree well with $N=2$ extrapolation.
As $\Psi_4$ is expected to be recovered to high accuracy by CCE, this is a first indication that $N=2$ extrapolation provides accurate predictions for the GW signal at infinity.

\begin{figure}
\includegraphics[width=0.95\columnwidth]{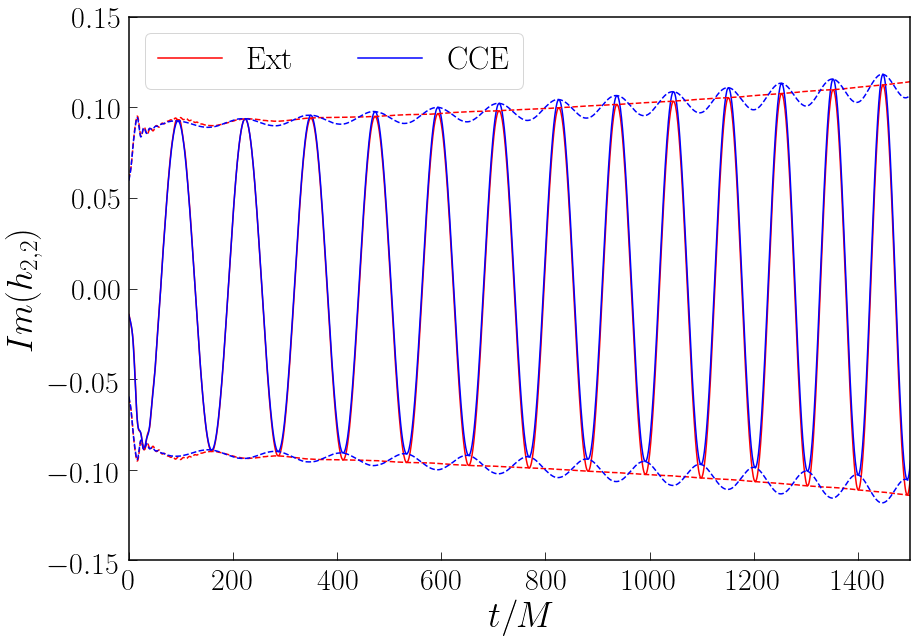}\\
  \caption{Imaginary part of the (2,2) mode of the GW strain ($h_\times$ polarization) using CCE extraction from $R=400M$ and extrapolation with $N=3$. The dotted line show the amplitude of the signal ($\sqrt{|h|^2}$). In the CCE results, the average
  value of the strain is shifted away from $0$ during the first $\Delta t\sim 1000M$.}
  \label{fig:CceDrift}
\end{figure}

Gravitational wave detectors, however, measure a projection of the complex strain $h=h_+ + i h_\times$, not $\Psi_4$. Unfortunately, CCE predictions for $h$ suffer from a small drift in the average value of the strain over the first $1000M$ of evolution,
illustrated in Fig.~\ref{fig:CceDrift}. In Fig.~\ref{fig:CceDrift}, the average value of $h_\times$ increases, eventually leading to a constant shift between the CCE and extrapolated results for $t>1000M$, and 
visible oscillations in the inferred amplitude and phase of the GW signal.

This issue remains under investigation, although the likely cause of the early slow drift and late constant shift in the CCE strain is incomplete CCE initial data and a corresponding maladapted BMS frame. The evolution performed by CCE on outgoing null slices has a similar, but somewhat less dramatic, initial-data problem as the central Cauchy simulation: the accumulated effect of the past inspiral is difficult to estimate, and therefore transient effects appear during the early stage of the simulation. These initial data transients occur on a longer timescale than the Cauchy junk, due to the larger characteristic scale of the system. The CCE scale is set by the extraction radius, rather than the orbital scale of the compact merger. Finally, the initial transient leaves a lasting imprint on the CCE strain as an erroneous `memory' contribution.

\begin{figure}
\includegraphics[width=0.95\columnwidth]{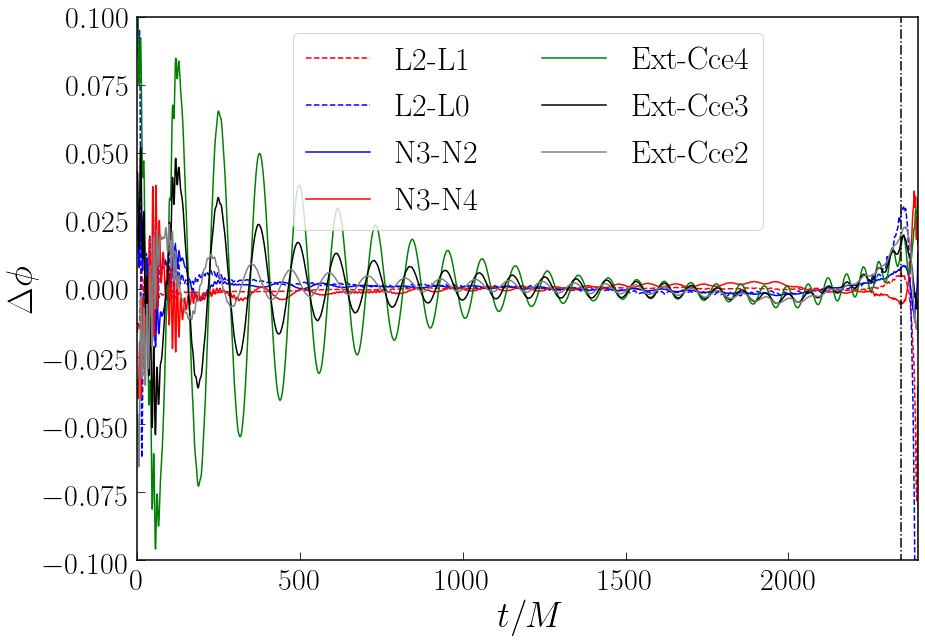}\\
\includegraphics[width=0.95\columnwidth]{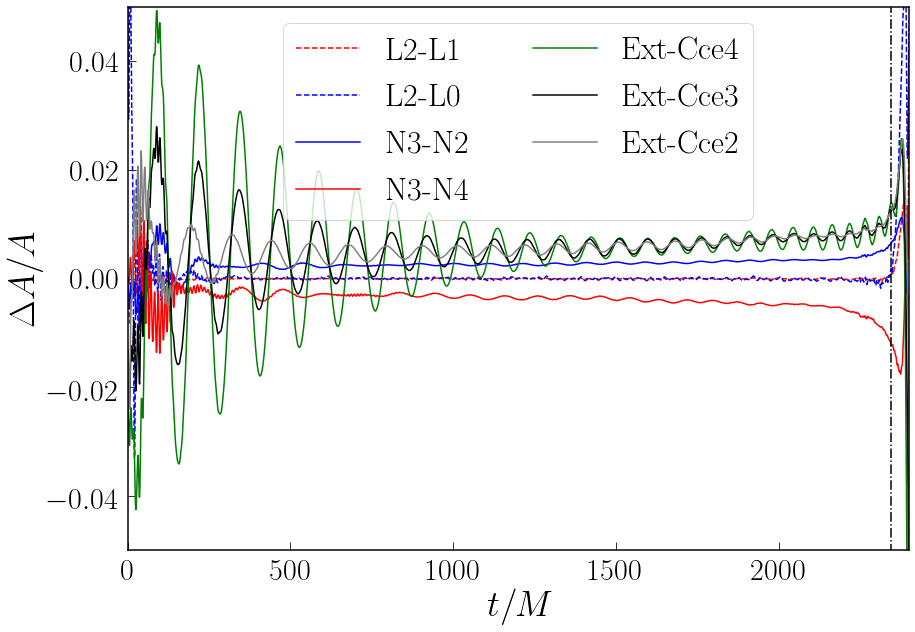}
  \caption{
 Same as Fig.~\ref{fig:CceVsExtPsi4Q3}, but for the $(2,2)$ mode of the GW strain.  
 For all CCE waveforms, we subtract a constant value from the strain (see text). 
  Early-time oscillation in the CCE results
  are due to the initial drift in the average value of the GW strain.}
  \label{fig:CceVsExtH}
\end{figure}

\begin{figure}
\includegraphics[width=0.95\columnwidth]{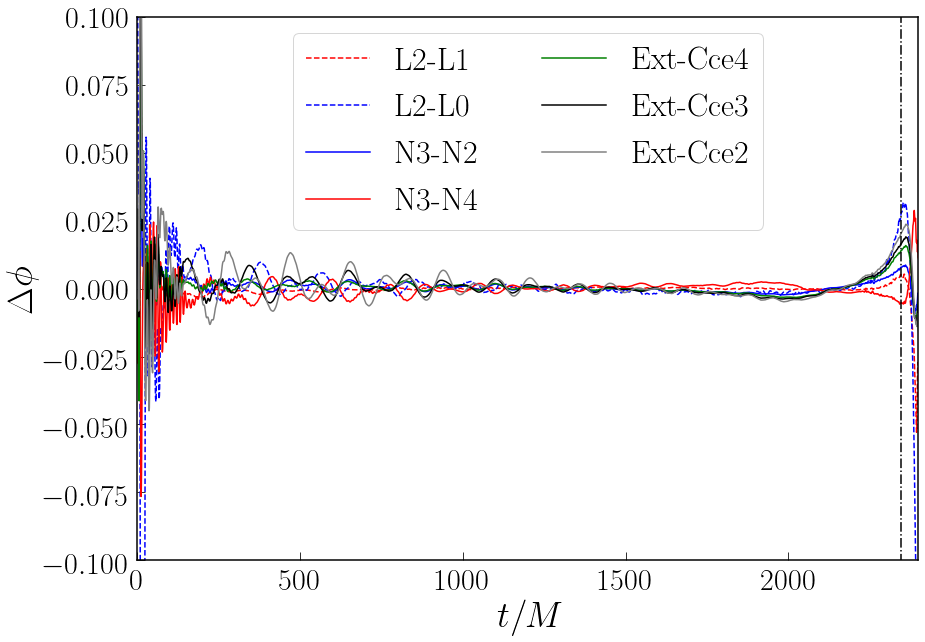}\\
\includegraphics[width=0.95\columnwidth]{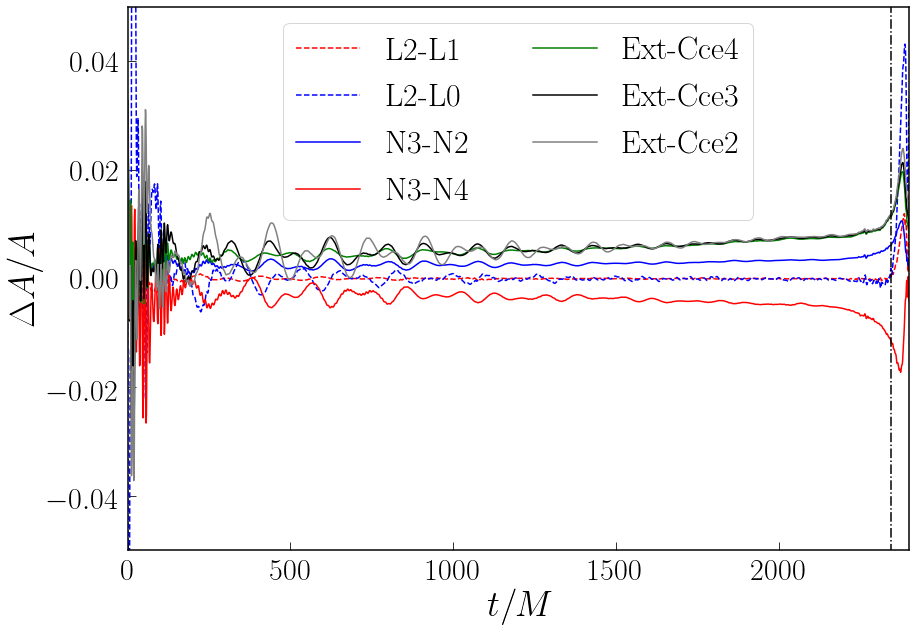}
  \caption{
 Same as Fig.~\ref{fig:CceVsExtH}, but after applying a highpass filter to all signals to remove the
 drift in the average value of the strain ($10^{\rm th}$ order Butterworth filter with critical frequency of $0.002M^{-1}\approx 72\,{\rm Hz}$).
 We note that this filter leads to changes in the phase of the signal that are larger than any of the errors displayed here; this plot indicates that most of the error observer
 in Fig.~\ref{fig:CceVsExtH} is indeed due to a slow drift in the average value of the strain, but the filtering does not provide us with a better waveform template.}
  \label{fig:CceVsExtH_Filter}
\end{figure}

This issue can be partially negated by applying a constant offset to $h_+,h_\times$, chosen to zero the average of the strain over a given time interval. 
Fig.~\ref{fig:CceVsExtH} shows phase and amplitude differences for the strain for case Q3S9, with the averaging performed over 8 GW cycles immediately following $t=1000M$. Large oscillations due to the drift in the strain dominate
the errors, but once we make abstraction of this issue, results are similar to what we obtained for $\Psi_4$: the various CCE waveforms are very consistent with each other, and agree well with $N=2$ extrapolation. The oscillations observed at the frequency of the GW signal in Fig.~\ref{fig:CceVsExtH} can be avoided by applying a highpass filter on all compared signals, as shown in Fig.~\ref{fig:CceVsExtH_Filter}. While this confirms the origin of the phase error, this filtered signal cannot be substituted for the original signal when performing model comparisons; in our attempts to filter the signal, no filter could remove the phase error due to the drift in the average value of the strain without introducing larger phase differences as a result of the filtering itself.

\begin{figure}
\includegraphics[width=0.95\columnwidth]{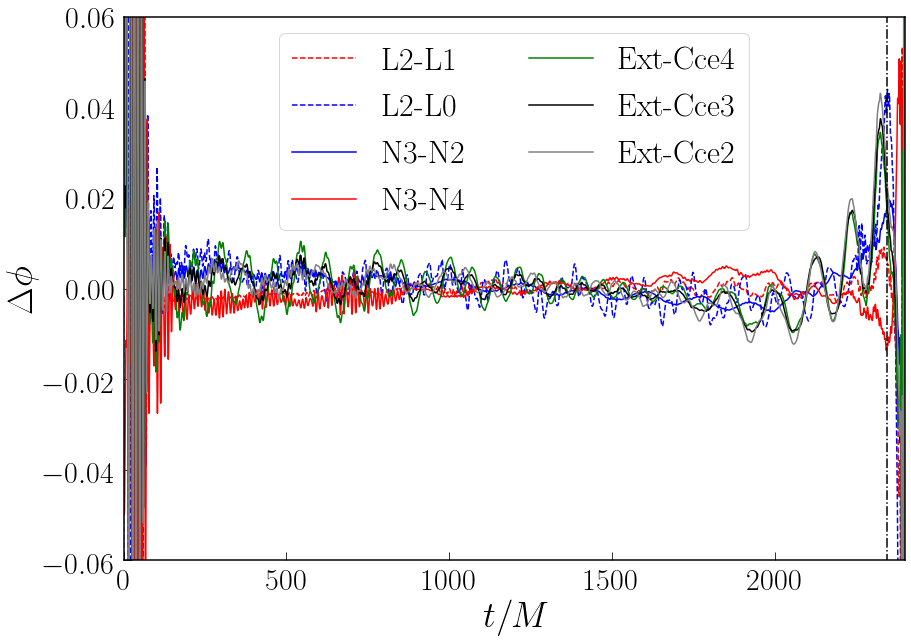}\\
\includegraphics[width=0.95\columnwidth]{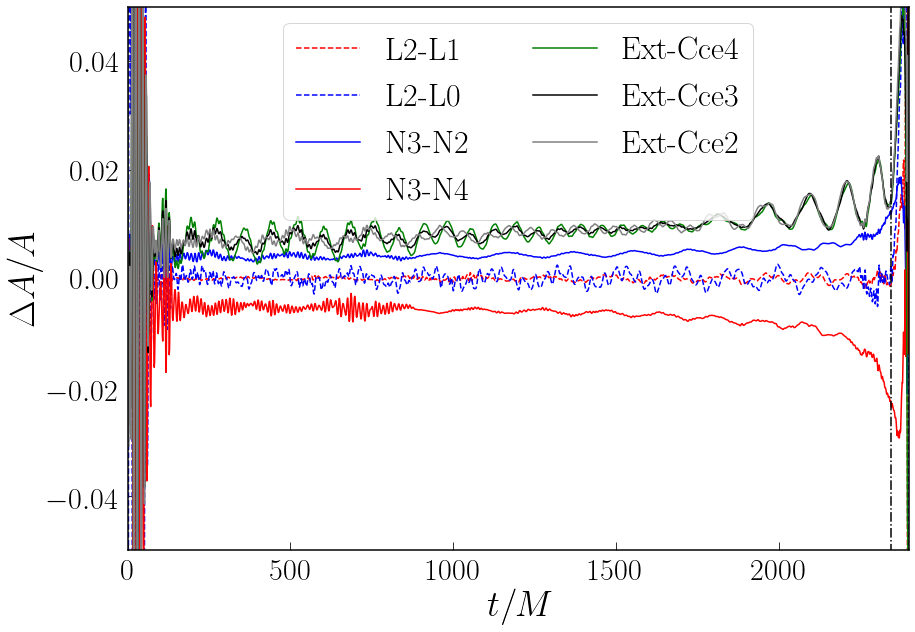}
  \caption{Same as Fig.~\ref{fig:CceVsExtH}, but for the $(3,3)$ mode of the GW strain. The CCE method provides much better result for this mode, with great self-consistency between CCE results
  using different extraction radii, and small differences between CCE and $N=2$ extrapolation.}
    \label{fig:CceVsExtH-33}
\end{figure}

Finally, we can follow the same procedure, but for the $(3,3)$ mode of the strain. Fig.~\ref{fig:CceVsExtH-33} shows the resulting differences between extrapolated and CCE waveforms. Higher-order
modes have very low amplitude early in the evolution of the binary, when the drift in the CCE strain occurs, and this appears to mitigate issues with the CCE method. 
As for $\Psi_4$ and the dominant mode of the strain, we find good agreement between all CCE waveforms and $N=2$ extrapolation.

\begin{figure*}
\includegraphics[width=0.95\textwidth]{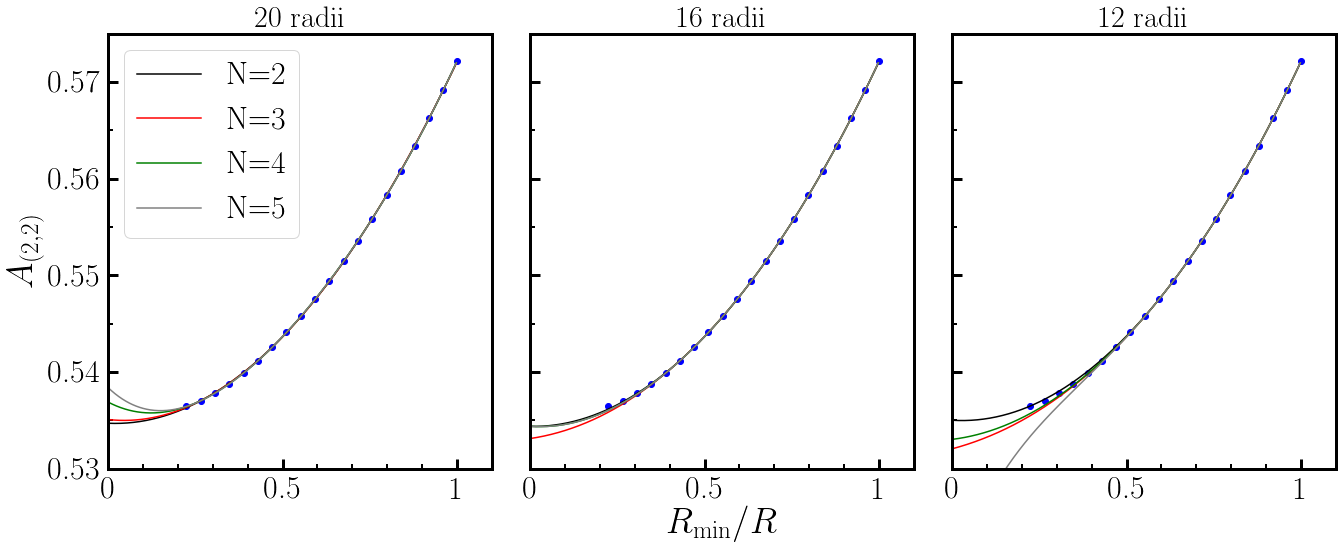}\\
  \caption{Amplitude of the $(2,2)$ mode of the GW strain at a retarded time $t=3000M$ for simulation Q3S9-L2. We show the values estimated at finite radii (normalized to $R_{\rm min}=100M$), and
  the extrapolation polynomials of order $N=2,3,4,5$. From left to right, we show extrapolation functions fitted to data at all 20 radii used by our simulation, the 16 smallest radii of that set, and the 
  12 smallest radii of that set (only the $N=3$ results is distinguishable from the others on the middle panel). While low-order extrapolation provides consistent and visually reasonable results, higher order extrapolation is less reliable. This can be contrasted with results
  in vacuum simulations, where results converge to a well-defined answer as $N$ increases.}
  \label{fig:ExtVsNR}
\end{figure*}

We can also gain some confidence in the accuracy of $N=2$ extrapolation by studying how robust extrapolation results are when the finite radii used by the fits change. This is illustrated in Fig.~\ref{fig:ExtVsNR},
where we show the results of $N=2,3,4,5$ extrapolation at a given retarded time ($t=3000M$) using all 20 finite radii, dropping the 4 radii farthest from the binary, and dropping the 8 radii farthest from the binary.
High-order extrapolation appears to be significantly impacted by noise in the finite radius measurements, leading to visibly problematic extrapolation functions (e.g. non-monotonous behavior of $A(r)$). On the other
hand, $N=2$ extrapolation provides very consistent results. While we find that the exact behavior of the higher-order extrapolation methods depends on the choice of retarded time under consideration, the robustness of $N=2$
extrapolation does not.

As a result of this exploration of CCE and waveform extrapolation, we can revisit our estimates for extrapolation errors, and recommendation for the `best' extrapolated waveform to use. The CCE waveforms (evolved from 3 different 
radii) and $N=2$ extrapolation provide results consistent to better than $1\%$ in the amplitude of the waveform, and better than $0.01$ radian in its phase {\it after alignment of the waveforms using an appropriate time and phase shift, 
and ignoring oscillations due to the drift in the average value of the strain in CCE}. For applications where waveform alignment is necessary (e.g. comparisons with analytical models, hybridization,...), this should provide us with
appropriate estimates of the extrapolation error {\it as long as we use $N=2$ extrapolation, and not higher order methods}. Comparing $N=2$ and $N=3$ extrapolation provides a reasonable upper bound on the extrapolation error with $N=2$
if that comparison is performed over the entire duration of the simulation.

\subsection{Comparison with analytical models}

\begin{table*}
\caption{Faithfulness of analytical models to the numerical results, for the $+$ polarization of the gravitational wave signal.
For each case, we consider face-on observation ($F$ suffix) and edge-on observation ($E$) suffix. All faithfulnesses are computed
using the pyCBC library, as discussed in more detail in the text. We also provide the frequency $f_{low}$ used as a lower bound for the calculation of the faithfulness, and the SNR
of the part of the signal above $f_{\rm low}$ for systems at a distance of $100\,{\rm Mpc}$. All calculations are performed using the Zero-Detuned High-Power noise power spectrum of LIGO.
  }
\label{tab:F}
\begin{tabular}{c|c|c|c|c|c|c|}
{\rm Model} & Q3S9-F  & Q3S9-E & Q4S9-F &  Q4S9-E & Q3S75p-F & Q3S75p-E\\
\hline
$f_{low}$ [Hz] & 300 & 300 & 250 & 250 & 300 & 300\\
SNR [100Mpc] & 12.5 & 5.9 & 16.7 & 7.9 & 11.7 & 5.1\\
\hline
\hline
IMRPhenomXAS & 0.972 & 0.957 & 0.981 & 0.955 & 0.988 &0.917\\
IMRPhenomXP & 0.969 & 0.955 & 0.976 & 0.951 & 0.986 & 0.959\\
IMRPhenomXHM & 0.971 & 0.964 & 0.979 & 0.967 & 0.987 & 0.928\\
IMRPhenomXPHM & 0.968 & 0.961 & 0.974 & 0.962 & 0.987 & 0.964\\
\hline
SEOBNRv4 & 0.966 & 0.952 & 0.976 & 0.950 & 0.987 & 0.915\\
SEOBNRv4P & 0.966 & 0.952 & 0.976 & 0.951\ & 0.970 & 0.955 \\
SEOBNRv4HM & 0.966 & 0.956 & 0.976 & 0.946 & 0.987 & 0.929 \\
SEOBNRv4PHM & 0.966 & 0.957 & 0.976 & 0.962 & 0.971 & 0.932\\
\hline
IMRPhenomPv2\_NRTidalv2 & 0.985\ & 0.972 & 0.969 & 0.947 & 0.984 & 0.957\\
SEOBNRv4T & 0.977 & 0.963 & 0.983 & 0.956 & 0.976 & 0.872 \\
\hline
IMRPhenomNSBH & 0.991 & 0.977 & 0.994 & 0.969 & 0.992 & 0.904\\
SEOBNRv4\_ROM\_NRTidalv2\_NSBH & 0.988 & 0.973 & 0.991 & 0.966 &  0.992 & 0.900
\end{tabular}
\end{table*}

With these error estimates in mind, we now move to a short investigation of the agreement between our numerical waveforms and commonly used waveform models in data analysis.
We note than an in-depth study of modeling uncertainties goes beyond the scope of this paper. Here, we are mostly interested in providing a broad overview
of the impact that various modeling choices and the physical effects of precession, higher modes, tidal effects during inspiral, and tidal disruption have on the agreement between numerical and analytical waveforms. We consider the following models, where the string `IMRPhenom' indicates a model in the family of phenomenological inspiral-merger-ringdown BBH models in the frequency-domain~\cite{Ajith-Babak-Chen-etal:2007,Ajith:2008,Ajith2009,Santamaria:2010yb,Khan:2015jqa,Husa:2015iqa,Hannam:2013oca,Schmidt:2012rh, Schmidt:2014iyl}, and `SEOBNR' a model in the family of time-domain inspiral-merger-ringdown BBH models based on the Effective-One-Body formalism and calibrated to numerical simulations~\cite{Buonanno99,Buonanno00,Damour2007,DIN,Nagar:2016ayt,Pan2010hz,Bohe:2016gbl,Taracchini:2012,Taracchini:2013rva,Pan:2013rra,Pan:2013tva,Barausse:2009xi,Barausse:2009aa,Babak:2016tgq,Barausse:2011ys,Pan:2011gk}:
\begin{itemize}
\item IMRPhenomXAS~\cite{Pratten:2020fqn}: an aligned-spin BBH model that only includes the dominant $(2,\pm 2)$ modes of the strain 
\item IMRPhenomXP~\cite{Pratten:2020ceb}: a BBH model that only includes the dominant $(2,\pm 2)$ modes of the strain and accounts for the main features of precession 
\item IMRPhenomXHM~\cite{Garcia-Quiros:2020qpx}: an aligned-spin BBH model that includes higher-order modes
\item IMRPhenomXPHM~\cite{Pratten:2020ceb}: a BBH model that includes higher-order modes and the main features of precession
\item SEOBNRv4~\cite{Bohe:2016gbl}: an aligned-spin BBH model that uses only the dominant $(2,\pm 2)$ modes to determine the strain 
\item SEOBNRv4P~\cite{Ossokine:2020kjp}: a BBH model that includes precession by describing all six spin degrees of freedom throughout the BBH coalescence, but uses only the dominant $(2,\pm 2)$ modes to determine the strain 
\item SEOBNRv4HM~\cite{Cotesta:2018fcv}: an aligned-spin BBH model that includes higher-order modes
\item SEOBNRv4PHM~\cite{Ossokine:2020kjp}: a BBH model that includes higher-order modes and precession by describing all six spin degrees of freedom throughout the BBH coalescence
\item IMRPhenomPv2\_NRTidalv2 ~\cite{Dietrich:2019kaq}: a model whose BBH baseline includes the main features of precession that also incorporates tidal effects based on calibrating analytical results to numerical simulations of NSNS binaries~\cite{Dietrich:2017aum,Dietrich:2018uni,Dietrich:2018upm,Dietrich:2019kaq}, but not higher order modes. This model only describes the inspiral, with the signals tapered to zero at the NSNS merger frequency predicted by numerical simulations, and does not attempt to model the disruption of a neutron star by a black hole companion.
\item SEOBNRv4T~\cite{Steinhoff:2016rfi,Hinderer:2016eia}: an aligned-spin model that includes analytical descriptions of tidal effects, but not higher-order modes. This model only describes the inspiral, with the signals tapered to zero at the NSNS merger frequency predicted by numerical simulations, and does not attempt to model the disruption of a neutron star by a black hole companion.
\item IMRPhenomNSBH~\cite{Thompson:2020nei}: an aligned-spin model specifically designed for BHNS binaries: it includes both tidal effects from ~\cite{Dietrich:2019kaq} and the disruption of the neutron star by the black hole. The model does not include higher-order modes.
\item SEOBNRv4\_ROM\_NRTidalv2\_NSBH~\cite{Matas:2020wab}: an aligned-spin model specifically designed for BHNS binaries based on a reduced-order-model approximation to the frequency-domain BBH signals predicted by the SEOBNRv4 model: it includes both tidal effects from ~\cite{Dietrich:2019kaq} and the disruption of the neutron star by the black hole. The model does not include  higher-order modes.
\end{itemize}

These models are generally representative of the latest iteration of the IMRPhenom and SEOBNR models (for recent models within another family of effective one body models see e.g.~\cite{Nagar:2020pcj,Akcay:2018yyh,Nagar:2019wds}). To determine the agreement between a model and numerical simulation, we calculate the faithfulness
\beq
F(h_1,h_2) = \max_{t_c,\phi_0}\left(\frac{\langle h_1,h_2 \rangle}{\sqrt{\langle h_1,h_1\rangle\langle h_2,h_2\rangle}}\right)
\eeq
with
\beq
\langle h_1,h_2\rangle = 4 {\rm Re} \int_{f_{\rm low}}^{f_{\rm high}} df \frac{h_1^*(f) h_2(f)}{S_n(f)}  
\eeq
and $S_n(f)$ the one-sided power spectral density of the detector noise. Here, we take $f_{\rm max}$ to be very large ($\sim 100\,{\rm kHz}$), and set $f_{\rm low}$ to a value sufficiently large to avoid artifacts due to the finite length of the numerical simulations ($250\,{\rm Hz}$ for the $Q=4$ system, and $300\,{\rm Hz}$ for the $Q=3$ systems). The faithfulness is calculated using the pyCBC library~\cite{pyCBC}, and that same library is used to generate the waveform models. Table~\ref{tab:F} shows the faithfulness of the various analytical models to the highest-resolution numerical simulation (using $N=2$ extrapolation). We calculate $F$ for the $+$ polarization of the waveform and for observers located along the direction of the total angular momentum of the system (face-on) and in a direction orthogonal to the total angular momentum (edge-on). We note that the faithfulness of the low-resolution simulation to the high-resolution simulation is $F>0.9999$ for all configurations and orientations, much larger than the faithfulness of any of the models. 

For the precessing simulations, defining the initial spins require a few additional assumptions. For models that only include aligned spins, 
we define the black hole spin as the component of the spin aligned with the orbital angular momentum at the beginning of the numerical simulation. For models that do include precession, we also have to maximize $F$ over the phase of the precession of the spin at a reference frequency, or over the reference frequency at which the spin is defined (depending on the inputs of the model). 

We also calculate the SNR $\sqrt{<h,h>}$ of the numerical waveform above $f_{\rm low}$ for a binary located at $100\,{\rm Mpc}$, to put the faithfulness numbers into context. We note that this is not the SNR of the full BHNS waveform, as a large fraction of the SNR is at frequency $f<f_{\rm low}$. All faithfulness and SNR results are summarized in Fig.~\ref{tab:F}.

We can see some clear trend in these tabulated results. For systems observed face-on, the two models specifically designed for BHNS systems perform noticeably better, with $F \gtrsim 0.99$. Including tidal effects without accounting for the disruption of the neutron star helps for the Q3S9 system ($F\sim 0.98$ for the tidal models, $F\sim 0.97$ for the BBH models), but has no noticeable impact on the faithfulness for Q4S9 and Q3S75p. The use of higher-order modes and/or precession does not seem to impact $F$ when a system is observed face-on. The faithfulness is generally lower for systems observed edge-on rather than face-on. The IMRPhenomX BBH models also inidicates that for edge-on systems, $F$ improves when including higher-order modes and, for the precessing binary, when including precession. With the SEOB models, higher-order modes help us improve $F$ for the non-precessing systems, and including precession helps with the precessing system -- but the model that include both effects actually perform worse than the precession-only and higher-mode only models. Finally, going from Q3S9 to Q4S9 to Q3S75p, the impact of tides decreases while the impact of higher-order modes / precession increases. As a result, the non-precessing tidal models (SEOBNRv4T, IMRPhenomNSBH, SEOBNRv4\_ROM\_NRTidalv2\_NSBH) become less faithful, down to just $F\sim 0.90$ for Q3S75p. We note however that it is perfectly possible that an analytical waveform with parameters reasonably close to Q3S75p would provide a good match to the numerical results -- we do not here attempt to find the best matching waveform, but only look at the faithfulness of the waveform for fixed initial conditions.

\begin{figure}
\includegraphics[width=0.95\columnwidth]{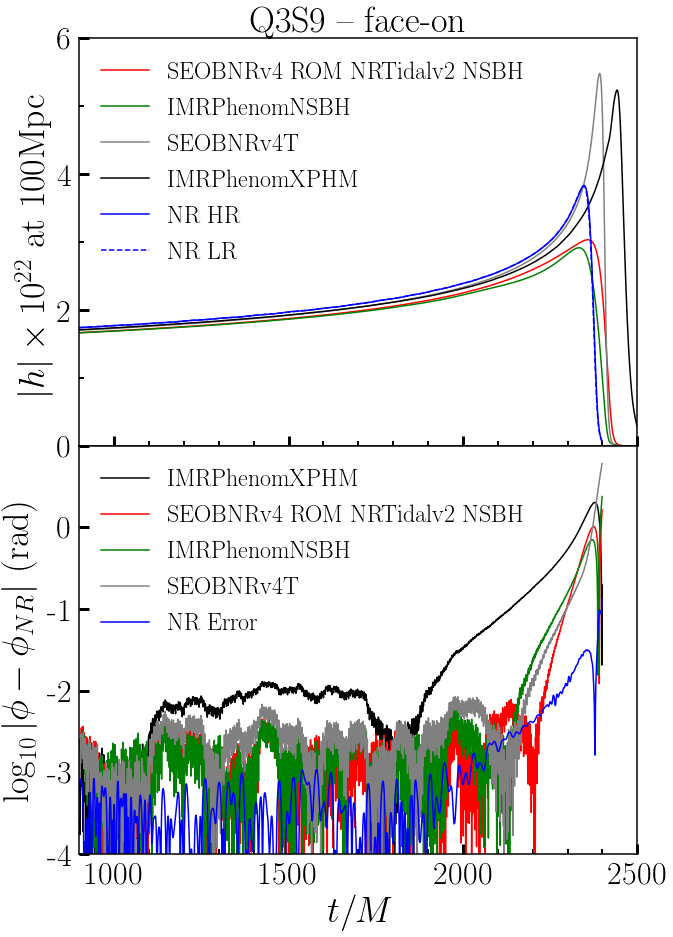}\\
  \caption{{\it Top}: Amplitude of the complex strain $h=h_++ih_\times$ for system Q3S9 seen face-on. We show results for our low- and high-resolution simulations, for the two BHNS models, as well as for one tidal model that does not include neutron star disruption and one BBH model that includes precession and higher-order modes.{\it Bottom}: Phase difference between the models and the high-resolution numerical result. All signals are aligned by minimizing the phase difference in the time interval $t\in [500,2000]M$.}
  \label{fig:Q3S9CompMod}
\end{figure}

\begin{figure}
\includegraphics[width=0.95\columnwidth]{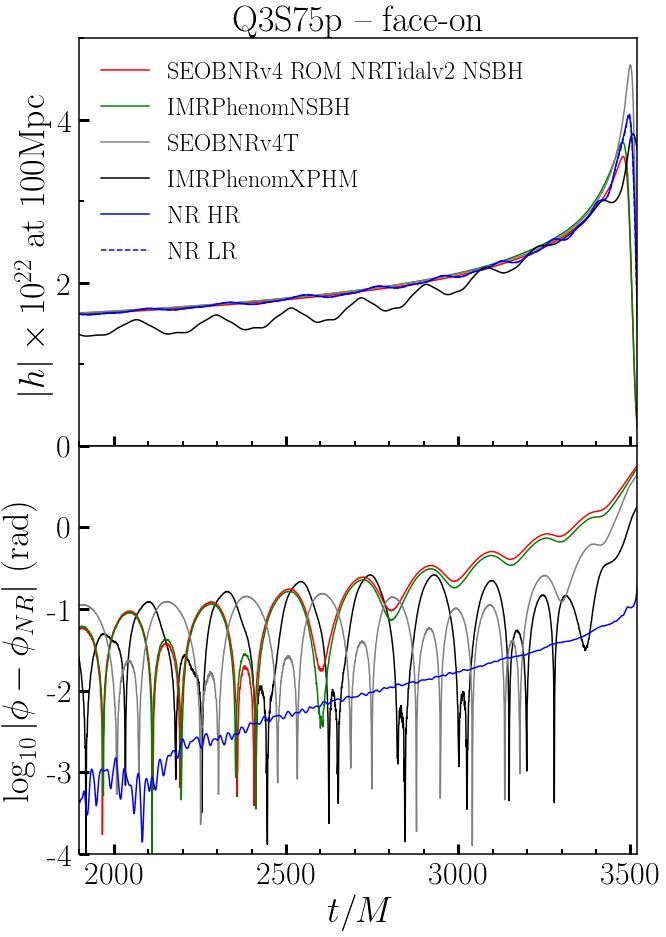}\\
  \caption{Same as Fig.~\ref{fig:Q3S9CompMod}, but for system Q3S75p.}
  \label{fig:Q3Sy5pCompMod}
\end{figure}

We can glean more information about the strengths and limits of various models by looking at Figs~\ref{fig:Q3S9CompMod}-\ref{fig:Q3Sy5pCompMod}. The amplitude plots best capture the impact of the models specifically designed for BHNS systems. The amplitude of the signal is very well captured for Q3S75p. For Q3S9, a system with a black hole spin higher than those used to calibrate BHNS models, the agreement is a little worse, but still noticeably better than for systems that do not account for the disruption of the neutron star. The phase errors for Q3S9 show the importance of including tidal phase corrections in the models. This is very different from our results for Q3S75p, for which the inclusion of precession and higher-order modes improves the phase accuracy far more than the inclusion of tidal effects.

 We also note that nearly identical results are obtained when comparing the second time derivative of the strain from analytical models to the CCE results for $\Psi_4$, indicating that one way to get around the drift in the strain when using CCE results to calibrate models could be to directly use $\Psi_4$ when performing these calibrations.

\section{Conclusions}

We presented a new set of long, high-accuracy BHNS waveforms generated using  the SpEC code, which are now publicly available. These waveforms sample regions of 
parameter space not covered by existing waveforms: high black hole spins, and one precessing system. All simulations are quite long by the standard of BHNS evolutions ($>13$ orbits),
and are of high accuracy in both phase ($0.2-0.4$ rad at merger) and amplitude ($\sim 1$\% errors). They should thus be particularly helpful for testing and calibrating future BHNS waveform models.

In previous BHNS simulations, errors due to the extrapolation of the waveform to infinity were typically negligible even when using conservative error estimates. We find that this is no longer the case
with our latest simulations. Accordingly, we perform a more careful study of extrapolation errors by comparing waveform extrapolation to CCE. Results for $\Psi_4$ at null infinity indicate very good agreement
between CCE initialized from different simulation radii, and between CCE and low-order (quadratic) extrapolation. Similar results appear to hold for the strain $h$, up to a drift in the time-averaged value of $h$
that appears when using CCE. Higher-order extrapolation appears less reliable. Overall, we conclude that while extrapolation errors remain small (compared to finite-resolution errors) in our simulations when using
quadratic extrapolation, one should avoid the use of higher-order methods in BHNS SpEC simulations. More practically, we thus recommend users of our waveform catalogue to take the highest resolution simulation
with quadratic extrapolation ($N=2$) as our `best' waveform when multiple resolutions and/or extrapolation orders are available.

Finally, we compute the faithfulness to our simulation of a range of existing binary black hole models, binary neutron star models, and BHNS models. We focus on the high frequency portion of the signal that can be studied directly with our numerical waveforms. We find that for face-on observations, two recent BHNS models perform quite well -- as already demonstrated in~\cite{Matas:2020wab} for our non-precessing systems. The inclusion of higher-order modes and/or precession effects is less crucial to high faithfulness at high frequency. A more careful study of phase errors however indicates that even for observations along the total angular momentum of the system, the inclusion of precessional effects can help reduce phase differences with respect to our precessing system. For non-precessing systems observed edge-on, higher-order modes and tidal effects are both significant. Finally, for the precessing system observed edge-on, only a few of the models used here manage to reach faithfulness $\gtrsim 0.95$ (all of them precessing models), and no model reaches faithfulness $>0.97$. Thus, there certainly remain important improvements that could be made to BHNS models by combining recent progress in the modeling of finite size effects with state-of-the-art results for precession in black hole binaries.

\begin{acknowledgments}
The authors thank Geert Raaijmakers and Andrew Matas for their help with the use of pyCBC.
UNH authors gratefully acknowledges
support from the NSF through grant PHY-1806278,
from the DOE through grant DE-SC0020435,
and from NASA through grant 80NSSC18K0565.
M.D gratefully acknowledges
support from the NSF through grant PHY-1806207.
H.P. gratefully acknowledges support from the
NSERC Canada. L.K. acknowledges support from NSF grant
PHY-1912081 and OAC-1931280. F.H. and M.S. acknowledge support from NSF Grants
PHY-170212 and PHY-1708213. F.H., L.K., N.D. and M.S. also thank
the Sherman Fairchild Foundation for their support. T.H. acknowledges support from the DeltaITP and NWO Projectruimte grant GW-EM NS.

\end{acknowledgments}

\bibliography{References/References}

\end{document}